\documentclass[12pt]{article}
\usepackage{booktabs,multirow}
\usepackage{apacite}
\usepackage{natbib}
\usepackage[hyphens,spaces]{url}
\usepackage{graphicx}
\usepackage{tabularx}
\usepackage{scalerel}
\usepackage{amsmath, amssymb}
\usepackage{blindtext}
\usepackage{subcaption}
\usepackage{caption}
\usepackage{booktabs}
\usepackage{anysize} 
\usepackage{lscape}
\usepackage{colortbl,hhline}
\usepackage{hyperref}
\usepackage{times}
\usepackage{setspace}
\usepackage{chngcntr}
\numberwithin{figure}{section}
\numberwithin{equation}{section}
\counterwithin{table}{section}

\newenvironment{sciabstract}{%
\begin{quote} \bf}
{\end{quote}}
\title{Multipopulation mortality modelling and forecasting: The multivariate functional principal component with time weightings approaches}

\author
{Ka Kin Lam\footnote{Corresponding author. E-mail: kkl21@leicester.ac.uk (K.K.Lam).},  Bo Wang\\
\\
\normalsize{School of Mathematics and Actuarial Science, University of Leicester,}\\
\normalsize{Leicester, LE1 7RH, UK}\\
}

\date{}
\begin{document} 
\baselineskip24pt
\maketitle 
\begin{sciabstract}
\begin{center}
Abstract
\end{center}
\normalfont Human mortality patterns and trajectories in closely related populations are likely linked together and share similarities. It is always desirable to model them simultaneously while taking their heterogeneity into account. This paper introduces two new models for joint mortality modelling and forecasting multiple subpopulations in adaptations of the multivariate functional principal component analysis techniques. The first model extends the independent functional data model to a multi-population modelling setting. In the second one, we propose a novel multivariate functional principal component method for coherent modelling. Its design primarily fulfils the idea that when several subpopulation groups have similar socio-economic conditions or common biological characteristics such close connections are expected to evolve in a non-diverging fashion. We demonstrate the proposed methods by using sex-specific mortality data. Their forecast performances are further compared with several existing models, including the independent functional data model and the Product-Ratio model, through comparisons with mortality data of ten developed countries. Our experiment results show that the first proposed model maintains a comparable forecast ability with the existing methods. In contrast, the second proposed model outperforms the first model as well as the current models in terms of forecast accuracy, in addition to several desirable properties.  
\end{sciabstract}

\section{Introduction}
There have been tremendous developments in the area of mortality modelling and forecasting over the last three decades. These include the pioneering mortality model proposed by \cite{lee1992modeling}, well-known as the Lee-Carter model. It rapidly gained credit and popularity, given its simplicity and ability to capture most variations in mortality pattern evolved over time. Several modifications and extensions of the Lee-Carter model have been put forward, see, for instance, \cite{lee2001evaluating}, \cite{booth2002applying}, \cite{renshaw2006cohort} and \cite{cairns2008modelling}. It is worth noting that \cite{hyndman2007robust} further extends the Lee-Carter model to a functional data framework, which includes non-parametric smoothing techniques, functional principal component decomposition and times series analysis to achieve the task of mortality modelling and forecasting. Although the models as mentioned earlier posed a great success in history, the single factor designs limit their capacity of mortality modelling and forecasting on solely one population. It seems improper to prepare a mortality forecast for an individual population in isolation from one another if they are closely linked together. For example, due to biological and behavioural reasons, male mortality rates have considerably been higher than female mortality rates, see \cite{kalben2000men}. However, if male mortality improvements are faster than female ones, but two genders are projected independently, the model may forecast male mortality rates lower than and eventually diverged further from female mortality rates. As such, it is always a significant challenge in human mortality modelling that the model can take multiple populations as well as their heterogeneity simultaneously into account. Several mortality models for multiple populations have been proposed in the literature over the last decade, see, for instance, \cite{delwarde2006application} and \cite{dowd2011gravity}. In more desirable cases, the model can further ensure that the forecasts for multiple related populations maintain certain structural relationships based on the extensive theoretical considerations and historical observations. A `coherent' or `non-divergent' model is one of the most well-suited cases in mortality modelling given the fact that the mortality of populations that are geographically close or otherwise related is driven by a common set of factors such as socio-economic, environmental and biological conditions and differences are unlikely to increase in the long run. Such coherent forecast models are also documented in the literature, see, for example, the earliest augmented common factor (ACF) model proposed by \cite{li2005coherent}, which is an extension of the Lee-Carter model with an additional common factor to capture both short-term divergence and long-term coherence among related populations. Variants and extensions of the ACF model have been subsequently developed, such as \cite{li2013poisson}, \cite{li2016multi} and \cite{chen2018sex}. Some others like the Age-Period-Cohort (APC) model proposed by \cite{cairns2011bayesian}, incorporate a mean-reverting stochastic process for two related populations and allow for different trends in mortality improvement rates in the short-run but parallel improvements in the long-run. The Product-Ratio model developed by \cite{hyndman2013coherent}, which models the product and ratio functions of the age-specific mortality rates of different populations individually through a functional principal component decomposition, achieves coherent mortality forecasts by constraining the forecast ratio function via a stationary time series model to appropriate constants. \cite{shang2016mortality} and \cite{wu2019coherent} use multilevel functional principal component analysis of aggregated and population-specific data to extract the common trend and population-specific residual trend among populations. The forecast of population-specific residual trend is restricted to be a stationary time process to achieve convergence in the long run. Some other developments in this field include \cite{jarner2011modelling}, \cite{hatzopoulos2013common} and \cite{wan2015swiss}. Also, see \cite{danesi2015forecasting} and \cite{enchev2017multi} for reviews and comparisons.
\par
In this paper, we propose two new models for mortality modelling and forecasting with the theoretical framework of multivariate functional principal component analysis techniques introduced by \cite{chiou2016multivariate} and \cite{happ2018multivariate}. The main objective of the multivariate functional principal component analysis is to model multiple sets of functional curves that may be correlated among others, which allows us to construct two new models on top of it. The first one is to model groups of population mortality rates with similarities across periods and ages together. The second model is a novel method for coherent mortality modelling and forecasting that captures the common trend and the population-specific trend of groups of mortality patterns and produces forecasts of different populations that do not diverge in the long run. The models are estimated using the weighted functional principal component algorithm \citep{hyndman2009forecasting}, which places more weights on recent mortality movements and so produces more realistic forecasts. 
\par More will be discussed in detail in the paper, and the rest of this article is organised as follows. In Section \ref{sec2: Theoretical background of FPCA}, we give a review of the theoretical background about univariate and multivariate functional principal component analyses. In Section \ref{sec3: Methodology}, we describe the general frameworks of two proposed multivariate functional principal component analysis models for mortality modelling and forecasting. We then illustrate the models by applying them to the sex-specific mortality rates for Japan with comparisons to two analogous functional data paradigms $-$ the independent FPCA model and the Product-Ratio model proposed by \cite{hyndman2007robust} and \cite{hyndman2013coherent}, in terms of the systematic differences and forecasting performances using sex-specific mortality data of ten developed countries in Section \ref{sec4: Applications}. We lastly conclude this article with discussions and remarks in Section \ref{sec5: Discussion and conclusion remarks}.

\section{Theoretical background of FPCA}\label{sec2: Theoretical background of FPCA}
Functional principal component analysis (FPCA) is the core technique applied primarily in this paper. It is a statistical method for analysing the variation of a bunch of functional curves in a dataset then reducing them from infinite dimensions to finite dimensions in principal component representations of variation \citep{ ramsay2007applied}. It can also be regarded as a functional extension of the multivariate PCA method, allowing the data dimension to increase from finite space to infinite space \citep{ramsay1991some}. After the Karhunen-Lo\`{e}ve theorem in expansions of a stochastic process proposed by \cite{karhunen1946spektraltheorie} and \cite{loeve1946fonctions}, the theoretical developments of FPCA can be divided into two main streams: the linear operator and the covariance operator perspectives, see, for example, \cite{besse1992pca}, \cite{ramsay2004functional}, \cite{yao2005functional}, \cite{hall2006properties} and \cite{bosq2012linear}. To get the readers well equipped with necessary concepts in this paper, we firstly give a brief review of univariate FPCA then move to discuss the theoretical framework of multivariate FPCA from the covariance operator point of view.   
\subsection{Univariate FPCA (UFPCA)}
Let $Y(x)$ be a mean  square  continuous ($L^{2}$-continuous) stochastic process on a domain $\mathcal{X}$ with a mean function $\mu{(x)} = \mathbb{E}(Y(x))$ and a covariance function $K(x,x') = \text{Cov}(Y(x), Y(x'))$ for all $x \in \mathcal{X}$. Assuming that there exists a covariance operator $\Lambda: L^{2}(\mathcal{X}) \rightarrow L^{2}(\mathcal{X})$ for any function $f \in L^{2}(\mathcal{X})$,
\[
\big(\Lambda f\big) (x) = \int_{\mathcal{X}} K(x,x') f(x') dx', \hspace*{0.5cm} \forall x \in \mathcal{X}.
\]
With the defined covariance operator $\Lambda$, we can perform a spectral analysis of the covariance function $K(x,x')$, such that
\[
\big(\Lambda\phi\big)(x) = \int_{\mathcal{X}} K(x,x')\phi(x')dx' = \lambda \phi(x),
\]
to obtain a set of orthonormal basis eigenfunctions $\{\phi_{n}(x)\}^{\infty}_{n=1}$ and a corresponding set of eigenvalues $\{\lambda_{n}\}^{\infty}_{n = 1}$, where $\lambda_{1} \geq \lambda_{2} \geq \cdots \geq 0$, representing the amount of variability in  $Y(x)$ explained by the $\{\phi_{n}(x)\}^{\infty}_{n=1}$. $Y(x)$ can now be represented as an infinite linear combination of the orthonormal functions by the Karhunen-Lo\`{e}ve theorem, that is
\[
Y(x) = \mu{(x)} + \sum_{n=1}^{\infty} \beta_{n}\phi_{n}(x),  \hspace*{0.5cm} \forall x \in \mathcal{X},
\]
where $\beta_{n}$ is the principal component score with
\[
\beta_{n} = \int_{\mathcal{X}} (Y(x) - \mu(x)) \phi_{n}(x)dx.
\]
The principal component scores $\{\beta_{n}\}_{n=1}^{\infty}$ are uncorrelated random variables with mean zero and variance $\{\lambda_{n}\}_{n=1}^{\infty}$. $\beta_{n}$ serves as the weight and the projection of the centred stochastic process $(Y(x) - \mu(x))$ in the direction of the $n$-th eigenfunction $\phi_{n}(x)$ in the Karhunen-Lo\`{e}ve representation of $Y(x)$. As the eigenvalues decrease quickly towards zero, only the first few eigenfunctions are needed to represent the most important features of $Y(x)$. The truncated Karhunen-Lo\`{e}ve expansion with the optimal first $N$-dimensional approximations to $Y(x)$ can be written as
\begin{equation}\label{eq: The truncated KL expansions}
Y(x) = \mu{(x)} + \sum_{n=1}^{N} \beta_{n}\phi_{n}(x), \hspace*{0.5cm} \forall x \in \mathcal{X},
\end{equation}
and thus reduce the infinite dimension of functional data into finite dimensions in principal direction of variation which is often used in practice for data analysis, e.g. for regression or for clustering \citep{ramsay2007applied}.   
 
\subsection{Multivariate FPCA (MFPCA)}
We now concern with multivariate functional data and the way to establish a direct link from the Karhunen-Lo\`{e}ve representation of univariate functional data discussed previously to the Karhunen-Lo\`{e}ve representation of multivariate functional data. Consider a random sample which consists of $p \geq 2$ sets of functions $Y^{(1)}(x),\cdots,Y^{(p)}(x)$ on a domain $\mathcal{X}$ for all  $x \in \mathcal{X}$. Combining all different sets of functions in a vector $\boldsymbol{Y}(x)$, we have
\[
\boldsymbol{Y}(x) = (Y^{(1)}(x),\cdots,Y^{(p)}(x))^{T} \in \mathbb{R}^{p}
\]
with a mean function 
\[
\boldsymbol{\mu}(x) =  (\mathbb{E}(Y^{(1)}(x)),\cdots,\mathbb{E}(Y^{(p)}(x)))^{T} =  (\mu^{(1)}(x),\cdots,\mu^{(p)}(x))^{T},
\]  
and a covariance function
\[
K_{ij}(x,x') = \text{Cov}(Y^{(i)}(x), Y^{(j)}(x')) = \mathbb{E}[(Y^{(i)}(x) - \mu^{(i)}(x))(Y^{(j)}(x') - \mu^{(j)}(x'))].
\]
Let a set of functions $\boldsymbol{f} = (f^{(1)},\cdots,f^{(p)})$ with an index for each $i = 1,\cdots,p:$ $f^{(i)} \in L^{2}(\mathcal{X})$. Assuming that there exists a covariance operator $\Gamma: L^{2}(\mathcal{X}) \rightarrow L^{2}(\mathcal{X})$ for all $f \in L^{2}(\mathcal{X})$ with the $i$-th element of $(\Gamma f)$,
\[
(\Gamma f)^{(i)}(x) = \sum_{j = 1}^{p} \int_{\mathcal{X}}K_{ij}(x,x') f^{(j)}(x') dx',\hspace*{0.5cm} \forall x \in \mathcal{X}.
\]
With the defined covariance operator $\Gamma$ and the similar structure in the univariate FPCA, we let $\big(\Gamma\psi\big)(x) = \nu \psi(x)$ by the spectral theorem on the covariance function $K_{ij}(x,x')$, where $\psi(x)$ is the orthonormal basis of eigenfunctions and $\nu$ is the corresponding eigenvalue. Then for all $i,j = 1,\cdots,p$ and $x \in \mathcal{X}$,  we have 
\[
\begin{aligned}
(\Gamma \psi)^{(i)}(x)
&= \sum_{j = 1}^{p} \int_{\mathcal{X}}K_{ij}(x,x')\psi^{(j)}(x')dx' = \nu \psi^{(i)}(x). 
\end{aligned}
\]
Without any loss of generality, we assume that each set of the functions $Y^{(1)}(x),\cdots,Y^{(p)}(x)$ has its finite univariate Karhunen-Lo\`{e}ve representation up to the  optimal first $N$-dimensional approximations as shown in Equation (\ref{eq: The truncated KL expansions}), i.e. $Y^{(i)}(x) = \mu^{(i)}{(x)} + \sum_{n=1}^{N} \beta^{(i)}_{n}\phi^{(i)}_{n}(x)$, and $K_{ij}(x,x')$ is a separable covariance function, i.e. $K_{ij}(x,x') = K_{i}(x,x') \cdot K_{j}(x,x')$. It holds for each $m,l = 1,\cdots,N$ and $i,j = 1,\cdots,p$ and for all $x\in \mathcal{X}$,
\begin{equation}\label{eq: Expansion of a separable covariance function}
\begin{aligned}[b]
(\Gamma \psi)^{(i)}(x) 
&= \sum_{j = 1}^{p} \sum_{m = 1}^{N} \sum_{l = 1}^{N}\int_{\mathcal{X}}\text{Cov}(\beta_{m}^{(i)}\phi_{m}^{(i)}(x), \beta_{l}^{(j)}\phi_{l}^{(j)}(x'))\psi^{(j)}(x')dx'\\
&= \sum_{j = 1}^{p} \sum_{m = 1}^{N} \sum_{l = 1}^{N}\text{Cov}(\beta_{m}^{(i)}, \beta_{l}^{(j)})\phi_{m}^{(i)}(x) \int_{\mathcal{X}} \phi_{l}^{(j)}(x')\psi^{(j)}(x')dx' = \nu \psi^{(i)}(x).
\end{aligned}
\end{equation}
For simplicity of notation, we denote $\text{Cov}(\beta_{m}^{(i)}, \beta_{l}^{(j)}) = Z_{ml}^{(ij)}$ and $\int_{\mathcal{X}} \phi_{l}^{(j)}(x')\psi^{(j)}(x')dx' = c_{l}^{(j)}$. Equation (\ref{eq: Expansion of a separable covariance function}) can be rewritten as  
\begin{equation}\label{eq: Covariance operator form with substitutions}
\sum_{j = 1}^{p} \sum_{m = 1}^{N} \sum_{l = 1}^{N}Z_{ml}^{(ij)}\phi_{m}^{(i)}(x)c_{l}^{(j)} = \nu \psi^{(i)}(x).
\end{equation}
Following a similar technique of \cite{zemyan2012classical} in solving Fredholm integral equations of the second kind with a separable covariance function, we can multiply  and integrate an orthonormal basis eigenfunction $\phi_{n}^{(i)}(x)$, for $n = 1,\cdots, N$, over the domain $\mathcal{X}$ for both sides of Equation (\ref{eq: Covariance operator form with substitutions}),
\[
\int_{\mathcal{X}}\phi_{n}^{(i)}(x) \cdot \sum_{j = 1}^{p} \sum_{m = 1}^{N} \sum_{l = 1}^{N}Z_{ml}^{(ij)}\phi_{m}^{(i)}(x)c_{l}^{(j)} dx = \nu \int_{\mathcal{X}}\phi_{n}^{(i)}(x) \cdot  \psi^{(i)}(x) dx.
\]
Due to the orthonormality, $\int_{\mathcal{X}} \phi_{n}^{(i)}(x) \cdot \phi_{m}^{(i)}(x)dx = 1$ if $m = n$, and $\int_{\mathcal{X}} \phi_{n}^{(i)}(x) \cdot \phi_{m}^{(i)}(x)dx = 0$ if $m\neq n$ otherwise. Denoting $\int_{\mathcal{X}}\phi_{n}^{(i)}(x) \cdot  \psi^{(i)}(x) dx = c_{n}^{(i)}$, it holds with the simplified notations
\begin{equation}\label{eq: Eigenequation in the discrete case}
\sum_{j = 1}^{p} \sum_{l = 1}^{N} Z_{nl}^{(ij)} c_{l}^{(j)} = \nu c_{n}^{(i)}.
\end{equation}
For a given value of $\nu$, the solvability of this linear system correlates
with the solvability of the integral equation. Since $i,j$ and $n,m,l$ are arbitrarily notated, Equation (\ref{eq: Eigenequation in the discrete case}) can be represented in matrix form when we consider it as a whole, which is equivalent to an eigenequation, i.e.
\begin{equation}\label{eq: Eigenequation in matrix form}
\boldsymbol{Zc} = {\nu}\boldsymbol{c},
\end{equation}
or 
\[
\begin{pmatrix} 
{\boldsymbol{Z}^{(11)}} & \hdots & \boldsymbol{Z}^{(1p)}\\ \vdots &  \ddots & \vdots \\ \boldsymbol{Z}^{(p1)} & \hdots & \boldsymbol{Z}^{(pp)}
\end{pmatrix} 
\begin{pmatrix} 
{\boldsymbol{c}^{(1)}} \\ \vdots \\ {\boldsymbol{c}^{(p)}}
\end{pmatrix} = 
{\nu}
\begin{pmatrix} 
{\boldsymbol{c}^{(1)}} \\ \vdots \\ {\boldsymbol{c}^{(p)}}
\end{pmatrix}
\]
with a positive semidefinite block matrix $\boldsymbol{Z}^{(ij)} \in \mathbb{R}^{N \times N}$ and an eigenvector $\boldsymbol{c}^{(i)} \in \mathbb{R}^{N}$ entries. With a matrix eigenanalysis performed on Equation (\ref{eq: Eigenequation in matrix form}), we can obtain a set of orthonormal eigenvectors  $\{\boldsymbol{c_{n}}\}^{N}_{n=1}$ of $\boldsymbol{Z}$, corresponding a set of eigenvalues $\{\nu_{n}\}^{N}_{n = 1}$, where $\nu_{1} \geq \cdots \geq \nu_{N} \geq 0$.
\par Substituting the orthonormal eigenvector $\boldsymbol{c_{n}}$ and the corresponding eigenvalue $\nu_{n}$ into Equation (\ref{eq: Covariance operator form with substitutions}), the eigenfunction $\psi^{(i)}_{n}(x)$ of $\Gamma$ is given by their elements:
\[
\psi_{n}^{(i)}(x) = \frac{1}{\nu_{n}} \sum_{j = 1}^{p} \sum_{m = 1}^{N} \sum_{l = 1}^{N}Z_{ml}^{(ij)}[\boldsymbol{c_{n}}]_{l}^{(j)} \phi_{m}^{(i)}(x) =   \sum_{m = 1}^{N}[\boldsymbol{c_{n}}]_{m}^{(i)}\phi_{m}^{(i)}(x),  \hspace*{0.5cm} \forall x \in \mathcal{X},
\]
where $[\boldsymbol{c_{n}}]^{(i)}$ denotes the $i$-th block of the orthonormal eigenvector  $\boldsymbol{c_{n}}$ of $\boldsymbol{Z}$ corresponding to its eigenvalue $\nu_{n}$. The truncated multivariate Karhunen-Lo\`{e}ve expansions with the first optimal $N$-dimensional approximations to $Y^{(i)}(x)$ can be written as
\begin{equation}
Y^{(i)}(x) = \mu^{(i)}{(x)} + \sum_{n=1}^{N} \rho_{n}\psi^{(i)}_{n}(x), \hspace*{0.5cm} \forall x \in \mathcal{X},
\end{equation}
where $\rho_{n}$ is the multivariate principal component score with
\[
\begin{aligned}
\rho_{n} 
&= \sum_{i = 1}^{p}  \int_{\mathcal{X}} (Y^{(i)}(x) - \mu^{(i)}(x)) \psi^{(i)}_{n}(x)dx\\
&= \sum_{i = 1}^{p}  \int_{\mathcal{X}} (Y^{(i)}(x) - \mu^{(i)}(x))\sum_{m = 1}^{N}[\boldsymbol{c_{n}}]_{m}^{(i)}\phi_{m}^{(i)}(x)\\
& = \sum_{i = 1}^{p} \sum_{m = 1}^{N}[\boldsymbol{c_{n}}]_{m}^{(i)}\int_{\mathcal{X}} (Y^{(i)}(x) - \mu^{(i)}(x))\phi_{m}^{(i)}(x) = \sum_{i = 1}^{p} \sum_{m=1}^{N}[\boldsymbol{c_{n}}]_{m}^{(i)}\beta^{(i)}_{m},
\end{aligned}
\]
where $\beta^{(i)}_{m}$  is the $m$-th univariate principal component score of the $i$-th element. \\
The mean and the covariance of $\rho_{n}$ can be derived for all $i,j = 1,\cdots,p$ and $x \in \mathcal{X}$ as
\[ \mathbb{E}(\rho_{n}) = \sum_{i = 1}^{p}  \int_{\mathcal{X}} \mathbb{E}(Y^{(i)}(x) - \mu^{(i)}(x))\psi^{(i)}_{n}(x)dx = 0,\] 
since $\mathbb{E}((Y^{(i)}(x) - \mu^{(i)}(x)) = 0$, and
\begin{equation}\label{eq: Covariance of multivariate functional principal component scores}
\begin{aligned}[b]
\text{Cov}(\rho_{n},\rho_{m}) &= \mathbb{E}\bigg( \sum_{i = 1}^{p}  \int_{\mathcal{X}}(Y^{(i)}(x) - \mu^{(i)}(x))\psi^{(i)}_{n}(x)dx \cdot \sum_{j = 1}^{p}\int_{\mathcal{X}}(Y^{(j)}(x') - \mu^{(j)}(x'))\psi^{(j)}_{m}(x')dx'\bigg)\\
&= \sum_{i = 1}^{p}\int_{\mathcal{X}} \sum_{j = 1}^{p} \int_{\mathcal{X}} K_{ij}(x,x') \psi^{(j)}_{m}(x')dx' \cdot \psi^{(i)}_{n}(x)dx \\
&= \sum_{i = 1}^{p} \int_{\mathcal{X}}  \nu_{m}\psi^{(i)}_{m}(x) \cdot \psi^{(i)}_{n}(x)dx\\
&= \nu_{m} \sum_{i = 1}^{p} \int_{\mathcal{X}} \psi^{(i)}_{m}(x) \cdot \psi^{(i)}_{n}(x)dx.
\end{aligned}
\end{equation}
Because of the orthonormality, $\int_{\mathcal{X}} \psi^{(i)}_{m}(x) \cdot \psi^{(i)}_{n}(x)dx = 1$ if $m = n$, and $\int_{\mathcal{X}} \psi^{(i)}_{m}(x) \cdot \psi^{(i)}_{n}(x)dx = 0$ if $m\neq n$ otherwise.
\subsection{Algorithm of estimating the MFPCA}
In practice, the natural path from the univariate to the multivariate functional principal component analysis discussed above can be summarised in a simple algorithm introduced by \cite{happ2018multivariate}. Given a random sample consists of $p \geq 2$ sets of functions $Y^{(1)}(x),\cdots,Y^{(p)}(x)$ on a domain $\mathcal{X}$ for all  $x \in \mathcal{X}$, the algorithm comprises the following four steps:  
\begin{enumerate}
	\item Perform a univariate functional principal component analysis for each element $Y^{(i)}(x)$ consisting of the observed curves $\{Y^{(i)}_{t}(x)\}_{t=1}^{T}$. This gives us a set of estimated principal component scores $\{\hat{\beta}^{(i)}_{t,n} \}^{N}_{n = 1}$ for each observation unit $t = 1 ,\cdots, T $ and estimated eigenfunctions $\{\hat{\phi}^{(i)}_{n}(x)\}^{N}_{n = 1}$ with the first suitably chosen $N$-dimensional approximations to each $Y^{(i)}(x)$. 
	\item Combine all the estimated principal component scores into a single large matrix $\boldsymbol{\Xi}$ where
	\[
	\boldsymbol{\Xi} = 
	\begin{pmatrix}
	{\hat{\beta}^{(1)}_{1,1}} & \hdots & \hat{\beta}^{(1)}_{1,N} & \hdots & {\hat{\beta}^{(p)}_{1,1}} & \hdots & \hat{\beta}^{(p)}_{1,N} \\ \vdots &  \ddots & \vdots &\ddots & \vdots &  \ddots & \vdots\\ \hat{\beta}^{(1)}_{T,1} & \hdots & \hat{\beta}^{(1)}_{T,N} & \hdots & {\hat{\beta}^{(p)}_{1,N}} & \hdots & \hat{\beta}^{(p)}_{T,N}
	\end{pmatrix}
	\in \mathbb{R}^{T \times pN}
	\]
	and estimate the joint covariance matrix $\hat{\boldsymbol{Z}}$ = $\frac{1}{N-1}\boldsymbol{\Xi}^{T}\boldsymbol{\Xi}$. 
	\item Perform a matrix eigenanalysis for $\hat{\boldsymbol{Z}}$ to obtain a set of  estimated orthonormal eigenvectors  $\{\hat{\boldsymbol{c}}_{n}\}^{N}_{n=1}$ and a set of corresponding eigenvalues $\{\hat{\nu}_{n}\}^{N}_{n = 1}$ of $\hat{\boldsymbol{Z}}$.
	
	\item Calculate the estimated multivariate eigenfunctions and the estimated multivariate principal component scores according to their $i$-th elements:
	\[
	\hat{\psi}_{n}^{(i)}(x) = \sum_{m = 1}^{N}[\boldsymbol{\hat{c}_{n}}]_{m}^{(i)}\hat{\phi}_{m}^{(i)}(x),\hspace*{0.5cm} \forall x \in \mathcal{X},
	\]
	and 
	\[
	\hat{\rho}_{t,n} = \sum_{i = 1}^{p} \sum_{m=1}^{N}[\boldsymbol{\hat{c}_{n}}]_{m}^{(i)}\hat{\beta}^{(i)}_{t,m}.
	\]
\end{enumerate}
The empirical truncated multivariate Karhunen-Lo\`{e}ve representation with the first optimal $N$-dimensional approximations to $Y_{t}^{(i)}(x)$ is
\[
\hat{Y}_{t}^{(i)}(x) = \hat{\mu}^{(i)}{(x)} + \sum_{n=1}^{N} \hat{\rho}_{t,n}\hat{\psi}^{(i)}_{n}(x),\hspace*{0.5cm} \forall x \in \mathcal{X},
\]
where $\hat{\mu}^{(i)}{(x)} = \frac{1}{T}\sum_{t=1}^{T}Y_{t}^{(i)}(x)$, and the estimated multivariate principal component score $\hat{\rho}_{t,n}$ gives the individual weight of each observation unit $t$ for its corresponding estimated multivariate eigenfunction $\hat{\psi}^{(i)}_{n}(x)$.
\section{Methodology}\label{sec3: Methodology}
\subsection{Weighted MFPCA model for multi-population mortality rates forecasting}
In this section we firstly introduce our new model, namely weighted MFPCA (wMFPCA) model for forecasting mortality rates of several subpopulations within a large population simultaneously.\par Let $Y^{(i)}_{t}(x)$ denote the log of the observed mortality rates of the $i$-th subpopulation for age $x$ in year $t$. We assume there is an underlying $L^{2}$-continuous smooth function $f^{(i)}_{t}(x)$ that we are observing with error and at discrete points of $x$. Our discrete observations are $\{ x_{j}, Y^{(i)}_{t}(x_{j})\}$, for $i = 1,\ldots,p, t = 1,\cdots,T, j = 1,\cdots,J$, such that
\[
Y^{(i)}_{t}(x_{j}) = f^{(i)}_{t}(x_{j}) + \sigma_{t}^{(i)}(x_{j})e^{(i)}_{t,j},
\] 
where $\{e^{(i)}_{t,j}\}_{t,j=1}^{T,J}$ are i.i.d. standard normal random variables, and $\sigma_{t}^{(i)}(x_{j})$ allows the amount of noise varying with age $x$.
\par In demographic modelling, it is often the case that more recent experience has greater relevance on the future behaviour than those data from the distant past. In view of this, we comprise a weighted functional component algorithm for the MFPCA model, allowing the forecasting results of the model to be based more on the recent data.
\par Let $\hat{f}^{(i)}_{t}(x)$ be a smoothed function estimated from the observation $Y^{(i)}_{t}(x_{j})$, and $w_{t} = \kappa( 1- \kappa)^{T - t}$ be a geometrically decaying weight with $0 < \kappa < 1$. The overall mean function $\mu^{(i)}(x)$ of $Y_{t}^{(i)}(x)$ is estimated by the weighted average
\[
\hat{{\mu}}^{(i)}(x) = \sum_{t = 1}^{T}w_{t}\hat{f}^{(i)}_{t}(x).
\]
The mean-centred functional data is denoted as $\hat{f}^{*(i)}_{t}(x) = \hat{f}^{(i)}_{t}(x) - \mu^{(i)}(x)$. We discretise $\hat{f}^{*(i)}_{t}(x)$ as a $T$ by $J$ matrix $\boldsymbol{\hat{F}^{(i)*}}$, then multiply $\boldsymbol{\hat{F}^{(i)*}}$ by $\boldsymbol{W}$, where $\boldsymbol{W} = \text{diag}(w_{1},\cdots,w_{T})$, such that $\boldsymbol{\hat{F}^{(i)}} = \boldsymbol{W}\boldsymbol{\hat{F}^{(i)*}}$. We then follow the algorithm of estimating MFPCA introduced in the previous section to calculate the weighted principal component scores and the weighted multivariate eigenfunctions using the functional form of $\boldsymbol{\hat{F}^{(i)}}$ to obtain $\hat{F}_{t}^{(i)}(x) =\sum_{n=1}^{N} \hat{\rho}_{t,n}\hat{\psi}^{(i)}_{n}(x)$ up to the first optimal $N$-dimensional approximations. We lastly combine the estimated weighted average with the estimated weighted multivariate eigenfunctions and the estimated multivariate weighted principal component scores to obtain the weighted MFPCA model for mortality modelling and forecasting of the $i$-th subpopulation with first optimal $N$-dimensional approximations, i.e.
\[
\hat{Y}_{t}^{(i)}(x) = \hat{\mu}^{(i)}{(x)} + \sum_{n=1}^{N} \hat{\rho}_{t,n}\hat{\psi}^{(i)}_{n}(x) + \hat{\sigma}_{t}^{(i)}(x)\hat{e}^{(i)}_{t}.
\]
\subsubsection{Out-of-sample forecasts and prediction intervals of the wMFPCA model}
Forecasts can be obtained by forecasting the weighted principal component scores $\{\hat{\rho}_{t,n}\}^{N}_{n=1}$ using time series models independently. There is no need to consider the vector autoregression (VAR) model for forecasting the weighted principal component scores as they are not correlated, see Equation (\ref{eq: Covariance of multivariate functional principal component scores}). $\{\hat{\rho}_{t,n}\}^{N}_{n=1}$ can be extrapolated using possibly non-stationary autoregressive integrated moving average (ARIMA) model, and we can select the order of the ARIMA model based on the Akaike information criterion (AIC) or the Bayesian information criterion (BIC).
\par Let $\hat{\rho}_{t+h,n}$ denote the $h$-step ahead forecast of $\hat{\rho}_{t,n}$, then the $h$-step ahead out-of-sample forecast of $\hat{Y}^{(i)}_{t}(x)$ is
\[
\hat{Y}^{(i)}_{t+h}(x) = \hat{\mu}^{(i)}{(x)} + \sum_{n=1}^{N} \hat{\rho}_{t+h,n}\hat{\psi}^{(i)}_{n}(x).
\]
We can also obtain the forecasting variance of the model by adding up the variances of all terms together given the fact that the components of the wMFPCA model are uncorrelated, such that
\[
\text{Var}(\hat{Y}^{(i)}_{t+h}(x)) = \hat{\tau}^{2}_{{\hat{\mu}}^{(i)}}(x) + \sum_{n=1}^{N}\hat{\nu}_{t+h,n}(\hat{\psi}^{(i)}_{n}(x))^{2} + (\hat{\sigma}^{(i)}_{t+h}(x))^{2},
\]
where $\hat{\tau}^{2}_{{\hat{\mu}}^{(i)}}(x)$ is the variance of the smoothed estimates of the mean function derived from the smoothing method applied, $\hat{\nu}_{t+h,n}$ is the estimated variance of $\hat{\rho}_{t+h,n}$ that can be obtained from the time series method used, and the estimated variance of forecast error $(\hat{\sigma}^{(i)}_{t+h}(x))^{2}$ is calculated by taking the average of observational variance from the historical data.
\par With the normality assumption on the model error and the known $\text{Var}(\hat{Y}^{(i)}_{t+h}(x))$, a $100(1-\alpha)\%$ prediction interval for $\hat{Y}^{(i)}_{t+h}(x)$ can be calculated as $\hat{Y}^{(i)}_{t+h}(x)\pm z_{\alpha}\sqrt{\text{Var}(\hat{Y}^{(i)}_{t+h}(x))}$, where $z_{\alpha}$ is the $(1-\alpha/2)$ quantile of the standard normal distribution.
\subsection{Coherent wMFPCA model for multi-population mortality rates forecasting}
We now introduce the idea of the coherent wMFPCA model, in the sense that the long term forecasts of several subpopulations within a large population will be non-divergent. \par Let $Y^{(i)}_{t}(x)$ be the log of the observed mortality rates of the $i$-th subpopulation for age $x$ in year $t$, $\{e^{(i)}_{t}\}_{t=1}^{T}$ are the i.i.d. standard normal random variables, and $\sigma_{t}^{(i)}(x)$ allows the amount of noise varying with age $x$. The coherent wMFPCA model has the following form:
\[
Y^{(i)}_{t}(x) = f^{(i)}_{t}(x)	+ \sigma_{t}^{(i)}(x)e^{(i)}_{t},\]
where
\[
f^{(i)}_{t}(x) = \mu(x) + \eta^{(i)}(x) + G_{t}(x) + Z_{t}^{(i)}(x).
\]
$f^{(i)}_{t}(x)$ is the smoothed mortality function of the $i$-th subpopulation for age $x$ in year $t$, $\mu(x)$ is the average of total mortality function, $\eta^{(i)}(x)$ is the mean of the $i$-th subpopulation specific deviation function from the averaged total mortality function, $G_{t}(x)$ is the common trend across all populations, and $Z_{t}^{(i)}(x)$ is the $i$-th subpopulation specific deviation trend.
\par In such a model, $\mu(x)$ and  $\eta^{(i)}(x)$ are unknown fixed functions, while $G_{t}(x)$ and $Z_{t}^{(i)}(x)$ are assumed to be independent zero mean stochastic processes to ensure identifiability \citep{shang2016mortality}, such that $G_{t}(x)$ and $Z_{t}^{(i)}(x)$ can then be decomposed by the (multivariate) Karhunen-Lo\`{e}ve representation as
\[
G_{t}(x) = \sum_{k=1}^{\infty}{\beta}_{t,k}{\phi}_{k}(x),
\]
\[
Z^{(i)}_{t}(x) = \sum_{l=1}^{\infty}{\gamma}_{t,l}{\varphi}^{(i)}_{l}(x),
\]
where $\{{\beta}_{t,k}\}^{\infty}_{k=1}$ and $\{{\phi}_{k}(x)\}^{\infty}_{k=1}$ are the corresponding principal component scores and the eigenfunctions of $G_{t}(x)$ while $\{{\gamma}_{t,l}\}^{\infty}_{l=1}$ and $\{{\varphi}^{(i)}_{l}(x)\}^{\infty}_{l=1}$ are the corresponding multivariate principal component scores and the multivariate eigenfunctions of $Z_{t}^{(i)}(x)$. It follows that $\{{\beta}_{t,k}\}^{\infty}_{k=1}$ is uncorrelated with $\{{\gamma}_{t,l}\}^{\infty}_{l=1}$. Following these expansions, the model can be expressed as
\[
f^{(i)}_{t}(x) = \mu(x) + \eta^{(i)}(x) + \sum_{k=1}^{\infty}{\beta}_{t,k}{\phi}_{k}(x) + \sum_{l=1}^{\infty}{\gamma}_{t,l}{\varphi}^{(i)}_{l}(x).
\]
\subsubsection{Estimation of the coherent wMFPCA model} 
\par We carry on the same weighted functional component algorithm applied in the wMFPCA model for the coherent wMFPCA model. The components of the coherent wMFPCA model can be obtained using the estimation procedures below in practice:
\begin{enumerate}
	\item Obtain the total mortality function among all subpopulations smoothed mortality functions, $\hat{g}_{t}(x) = \frac{1}{p}\sum_{i=1}^{p} \hat{f}^{(i)}_{t}(x)$, then calculate the weighted mean function of the total mortality function, $\hat{{\mu}}(x) = \sum_{t = 1}^{T}w_{t}\hat{g}_{t}(x)$, where $w_{t} = \kappa( 1- \kappa)^{T - t}$ is a geometrically decaying weight with $0 < \kappa < 1$. 
	\item Calculate the centered functional data $\hat{g}_{t}^{*}(x) = \hat{g}_{t}(x) - \hat{\mu}(x)$, then discretise $\hat{g}_{t}(x)$ as a $T$ by $J$ matrix $\boldsymbol{G^{*}}$, then multiply $\boldsymbol{G^{*}}$ by $\boldsymbol{W}$, where $\boldsymbol{W}= \text{diag}(w_{1},\cdots,w_{T})$, such that $\hat{\boldsymbol{G}} = \boldsymbol{W}\boldsymbol{G^{*}}$.
	\item Perform univariate FPCA on the functional form of $\hat{\boldsymbol{G}}$ to get $\hat{G}_{t}(x) =  \sum_{k=1}^{K}\hat{\beta}_{t,k}\hat{\phi}_{k}(x)$ up to the first optimal $K$-dimensional approximations. Let $\tilde{g}_{t}(x) = \hat{\mu}(x) + \sum_{k=1}^{K}\hat{\beta}_{t,k}\hat{\phi}_{k}(x)$ be the estimated weighted total mortality function.
	\item Calculate the deviation of the $i$-th subpopulation specific mortality function from the estimated weighted total mortality function, $\hat{d}_{t}^{(i)}(x) = \hat{f}^{(i)}_{t}(x) - \tilde{g}_{t}(x)$, then calculate the weighted mean of the subpopulation specific deviation function, $\hat{\eta}^{(i)}(x) = \sum_{t = 1}^{T}w_{t}\hat{d}_{t}^{(i)}(x)$.
	\item Obtain the demeaned functional data $\hat{z}_{t}^{(i)*}(x) = \hat{d}^{(i)}_{t}(x) - \hat{\eta}^{(i)}(x)$, then discretise $\hat{z}_{t}^{(i)*}(x)$ as a $T$ by $J$ matrix ${\boldsymbol{\hat{Z}^{(i)*}}}$, then multiply ${\boldsymbol{\hat{Z}^{(i)*}}}$ by $\boldsymbol{W}$, where $\boldsymbol{W} = \text{diag}(w_{1},\cdots,w_{T})$, to have ${\boldsymbol{\hat{Z}^{(i)}}} = \boldsymbol{W}{\boldsymbol{\hat{Z}^{(i)*}}}$.
	\item Perform multivariate FPCA on the functional form of ${\boldsymbol{\hat{Z}^{(i)}}}$ to obtain $\hat{Z}^{(i)}_{t}(x) =  \sum_{l=1}^{L}\hat{\gamma}_{t,l}\hat{\varphi}_{l}^{(i)}(x)$ up to the first optimal $L$-dimensional approximations.
\end{enumerate}
With all the estimated components obtained above, we can represent the coherent wMFPCA model for mortality modelling and forecasting of the $i$-th subpopulation, i.e.
\[
\hat{Y}^{(i)}_{t}(x) = \hat{\mu}(x) + \hat{\eta}^{(i)}(x) + \hat{G}_{t}(x) + \hat{Z}_{t}^{(i)}(x)	+ \hat{\sigma}_{t}^{(i)}(x)\hat{e}^{(i)}_{t},\]
or the full representation of the coherent wMFPCA model with the first optimal $K$-dimensional approximations to the common trend and the first optimal $L$-dimensional approximations to the $i$-th subpopulation deviation trend, such that
\[\hat{Y}^{(i)}_{t}(x) = \hat{\mu}^{(i)}(x) + \sum_{k=1}^{K}\hat{\beta}_{t,k}\hat{\phi}_{k}(x) + \sum_{l=1}^{L}\hat{\gamma}_{t,l}\hat{\varphi}_{l}^{(i)}(x) + \hat{\sigma}_{t}^{(i)}(x)\hat{e}^{(i)}_{t},\]
where $\hat{\mu}^{(i)}(x) = \hat{\mu}(x) + \hat{\eta}^{(i)}(x)$ is the mean function of the $i$-th subpopulation. 
\subsubsection{Out-of-sample forecasts and prediction intervals of the coherent wMFPCA model}
The $h$-step ahead out-of-sample forecast of $Y^{(i)}_{t}(x)$ can be represented as
\[\hat{Y}^{(i)}_{t+h}(x) = \hat{\mu}^{(i)}(x) + \sum_{k=1}^{K}\hat{\beta}_{t+h,k}\hat{\phi}_{k}(x) + \sum_{l=1}^{L}\hat{\gamma}_{t+h,l}\hat{\varphi}_{l}^{(i)}(x),
\]
where $\hat{\beta}_{t+h,k}$ and $\hat{\gamma}_{t+h,l}$ are the $h$-step ahead forecasts of the weighted principal component scores of the common trend and the $i$-th subpopulation specific deviation trend. $\hat{\beta}_{t+h,k}$ can be obtained using a univariate time series forecasting method, such as ARIMA model. To ensure the predictions of the subpopulations are coherent in the long term, the forecasts of all subpopulation deviation trends need to be restricted to be convergent and a stationary process, such that $\displaystyle{\lim_{h \to \infty}} \sum_{l=1}^{L} \big(\hat{\gamma}_{t+h,l}\hat{\varphi}_{l}^{(i)}(x) - \hat{\gamma}_{t+h,l}\hat{\varphi}_{l}^{(j)}(x) \big) = 0$. $\hat{\gamma}_{t+h,k}$ can hence be achieved using possibly any stationary autoregressive moving average (ARMA) process or autoregressive fractionally integrated moving average (ARFIMA) process. The order of the aforementioned time series models can be decided based on the Akaike information criterion (AIC) or the Bayesian information criterion (BIC).
\par Given the way that the coherent wMFPCA model has been constructed, each component is independent to the other components. Therefore, the forecast variance can be expressed by the sum of component variances, i.e.
\[
\text{Var}(\hat{Y}^{(i)}_{t+h}(x)) = \hat{\tau}^{2}_{{\hat{\mu}}^{(i)}}(x) + \sum_{k=1}^{K}\hat{u}_{t+h,k}(\hat{\phi}_{k}(x))^{2} + \sum_{l=1}^{L}\hat{\omega}_{t+h,l}(\hat{\varphi}^{(i)}_{l}(x))^{2} + (\hat{\sigma}^{(i)}_{t+h}(x))^{2},
\]
where $\hat{\tau}^{2}_{{\hat{\mu}}^{(i)}}(x)$ is the variance of the smoothed estimates of the mean function derived from the smoothing method used, $\hat{u}_{t+h,k}$ and $\hat{\omega}_{t+h,l}$ are the variances of $\hat{\beta}_{t+h,k}$ and $\hat{\gamma}_{t+h,l}$ that can be obtained from the time series methods applied, and the forecast error $(\hat{\sigma}^{(i)}_{t+h}(x))^{2}$ is the average of the observational variance estimated from the historical data.
\par With the normality assumption on the model error and the known $\text{Var}(\hat{Y}^{(i)}_{t+h}(x))$, a $100(1-\alpha)\%$ prediction interval for $\hat{Y}^{(i)}_{t+h}(x)$ can be calculated as $\hat{Y}^{(i)}_{t+h}(x)\pm z_{\alpha}\sqrt{\text{Var}(\hat{Y}^{(i)}_{t+h}(x))}$, where $z_{\alpha}$ is the $(1-\alpha/2)$ quantile of the standard normal distribution.
\par Note that the weights $\{w_{t}\}_{t=1}^{T}$ are controlled by the tuning parameter $\kappa$ in the geometrically decaying weighting approach embedded in the two proposed models. The larger $\kappa$ is, the faster the weights for the historical observations are decaying over time geometrically. In practice, the tuning parameter $\kappa$ can be determined by minimising the average root mean square error (RMSE) of all populations defined as 
\begin{equation}\label{eq: Root mean square error}
\text{RMSE} = \frac{1}{p}\sum_{i=1}^{p}\sqrt{\frac{1}{J}\sum_{j=1}^{J}\bigg(Y^{(i)}_{t+h}(x_{j}) - \hat{Y}^{(i)}_{t+h}(x_{j})\bigg)^{2}}.
\end{equation}
The value of the parameter $\kappa$ can alternatively be  specified as a $\mathit{prior}$, if there is a strong prior knowledge of how past data should be weighted \citep{wu2019coherent}. 
\par For selecting the optimal number of principal components in the two proposed models, we use a cumulative percentage of total variation method. We denote $N$ as a generic notation of the optimal number of principal components, and $N$ is determined by
\[
N = \underset{N:N\geq 1}{\text{argmin}}\bigg(\frac{\sum_{n=1}^{N}\hat{\lambda}_{n}}{\sum_{n=1}^{\infty}\hat{\lambda}_{n}} \geq P\bigg)
\]
where $\hat{\lambda}_{n}$ is the corresponding estimated eigenvalue of the principal components analysis, and $P$ $\geq$ 0.9 is set to be the minimum acceptance level as suggested by \cite{chiou2012dynamical}.
\section{Applications}\label{sec4: Applications}
In this section, we illustrate the two proposed models $-$ the wMFPCA model and the coherent wMFPCA model using sex-specific mortality data. We first present and plot the observed mortality dataset, then demonstrate the usefulness of these two models by forecasting of the sex-specific mortality rates of Japan. We show the forecasting results for males and females compared with the observed data. We further exhibit the ability of non-diverging long term forecasts of the proposed coherent wMFPCA model and finally assess the forecasting accuracy of the two proposed models in comparison to the Product-Ratio model and the independent FPCA model using the sex-specific mortality data of ten different developed countries. 
\subsection{Sex-specific mortality data of Japan}
\begin{figure}[!thb]
		\centering
	\begin{minipage}{0.78\textwidth}
		\centering
		\includegraphics[width=1\linewidth]{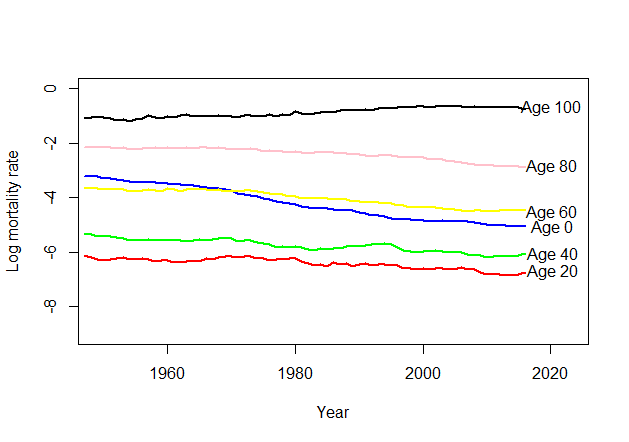}
		\subcaption{Male}
	\end{minipage}
	\begin{minipage}{0.78\textwidth}
		\centering
		\includegraphics[width=1\linewidth]{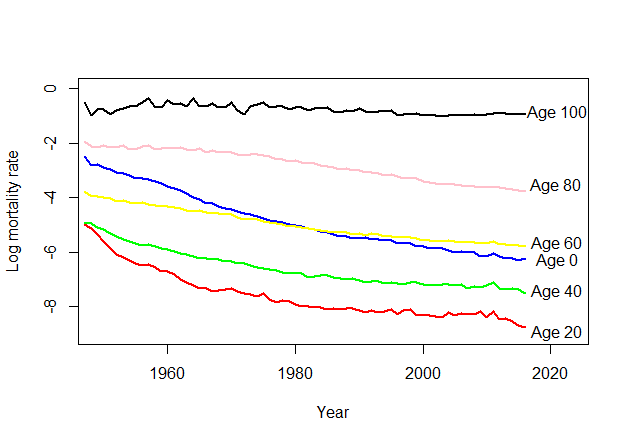}
		\subcaption{Female}
	\end{minipage}
	\caption{Log mortality rates for (a) male and (b) female from age 0 to age 100 with 20-year age intervals from the year 1947 to the year 2016 in Japan.}
	\label{fig: Log mortality rates for male and female from age 0 to age 100 with 20-year age interval}
\end{figure}

\begin{figure}[!thb]
	\centering
	\begin{minipage}{0.78\textwidth}
		\centering
		\includegraphics[width=1\linewidth]{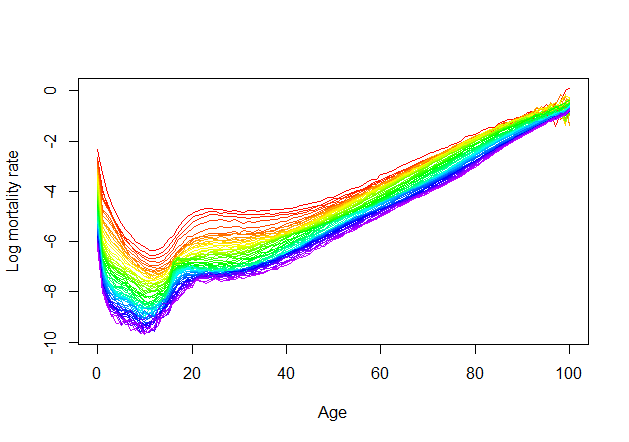}
		\subcaption{Observed}
	\end{minipage}
	\begin{minipage}{0.78\textwidth}
		\centering
		\includegraphics[width=1\linewidth]{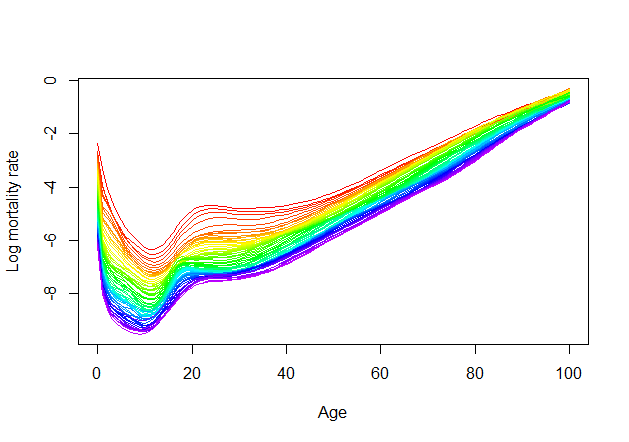}
		\subcaption{Smoothed}
	\end{minipage}
	\caption{(a) Observed and (b) smoothed log mortality rates for male from the year 1947 to the year 2016 in Japan, viewed as functional data curves with time-ordering indicated by the colours of the rainbow from red to violet.}
	\label{fig: Unsmoothed and smoothed log mortality rates for male}
\end{figure}

\begin{figure}[!thb]
	\centering
	\begin{minipage}{0.78\textwidth}
		\centering
		\includegraphics[width=1\linewidth]{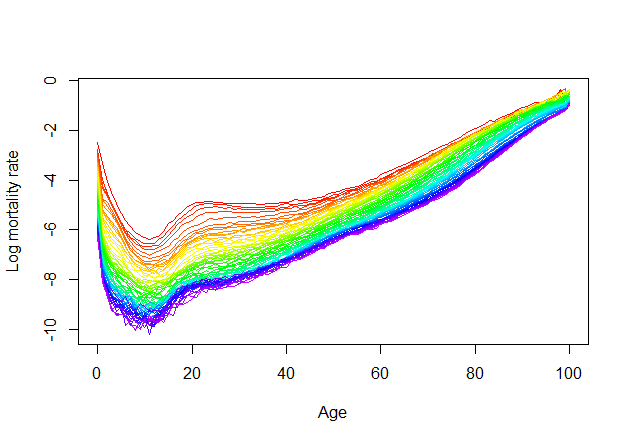}
		\subcaption{Observed}
	\end{minipage}
	\begin{minipage}{0.78\textwidth}
		\centering
		\includegraphics[width=1\linewidth]{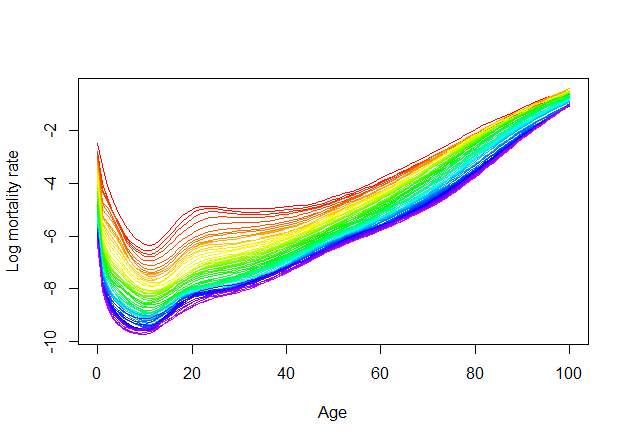}
		\subcaption{Smoothed}
	\end{minipage}
	\caption{(a) Observed and (b) smoothed log mortality rates for female from the year 1947 to the year 2016 in Japan, viewed as functional data curves with time-ordering indicated by the colours of the rainbow from red to violet.}
	\label{fig: Unsmoothed and smoothed log mortality rates for female}
\end{figure}

The sex-specific mortality rates data of Japan are available for the year 1947 to the year 2016 from the \cite{human2020university}. The database consists of central death rates by gender and single calendar year of age up to 110 years old. We restrict the data at the maximum age of 100 to avoid problems associated with erratic rates above age 100. The observed mortality rates curves are smoothed using penalised regression splines with a partial monotonic constraint so that each mortality curve is increasing above age 65 monotonically \citep{hyndman2007robust}. Figure \ref{fig: Log mortality rates for male and female from age 0 to age 100 with 20-year age interval} shows the log mortality rates for males and females for selected age groups as univariate time series. Figure \ref{fig: Unsmoothed and smoothed log mortality rates for male} and Figure \ref{fig: Unsmoothed and smoothed log mortality rates for female} present the same set of data as a batch of observed and smoothed curves (functional observation) in rainbow plots with time-ordering indicated by the colours of the rainbow, from red to violet. Figure \ref{fig: Unsmoothed and smoothed log mortality rates for male} and Figure \ref{fig: Unsmoothed and smoothed log mortality rates for female} both show that there are steady declines in male and female mortality rates at most ages over our examined period. The mortality curve patterns for male and female are reasonably similar, while for male, the mortality rates are generally higher than the mortality rates of female, particularly at around age 20. Despite the higher male mortality rates in comparison with female's, the mortality gap between male's and female's gets narrower over time at older ages.

\subsection{Sex-specific mortality forecasting by the wMFPCA model}
\begin{figure}[!thb]
	\begin{minipage}{1\textwidth}
		\centering
		\includegraphics[width=0.4\linewidth]{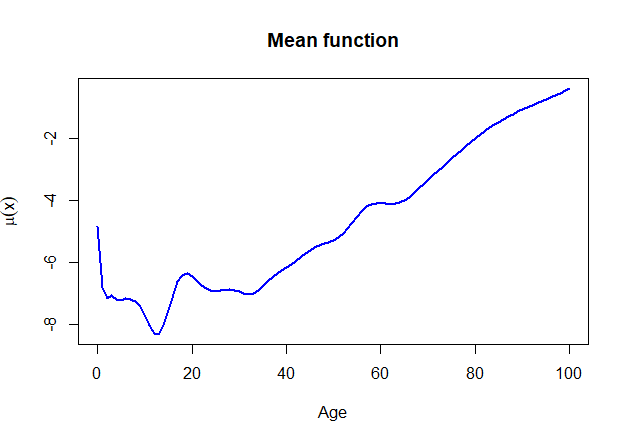}
		\includegraphics[width=0.4\linewidth]{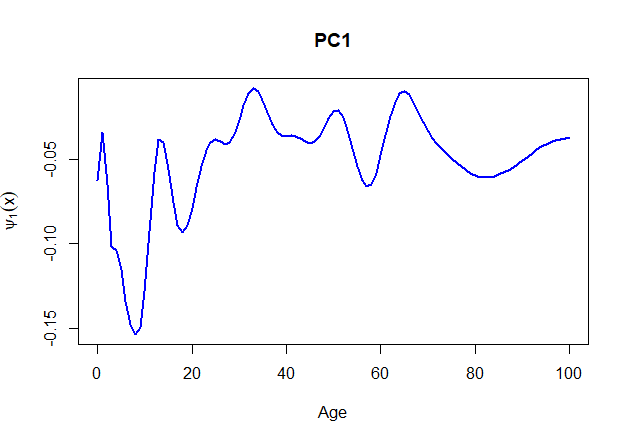}
	\end{minipage}
	\begin{minipage}{1\textwidth}
		\centering
		\includegraphics[width=0.4\linewidth]{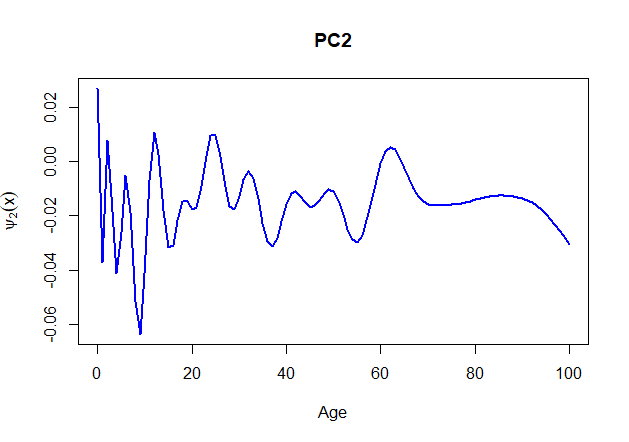}
		\includegraphics[width=0.4\linewidth]{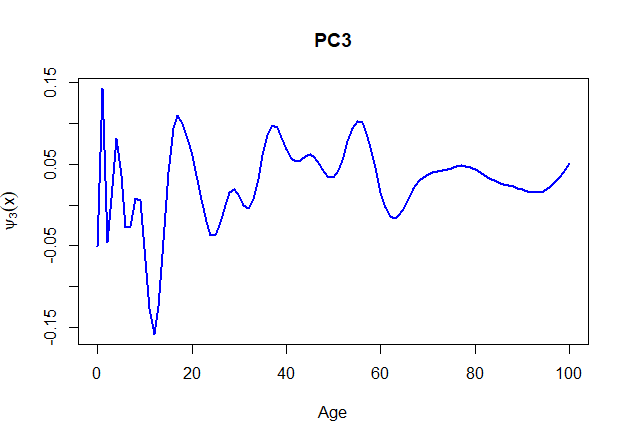}
		\subcaption{Male}
	\end{minipage}
	\begin{minipage}{1\textwidth}
		\centering
		\includegraphics[width=0.4\linewidth]{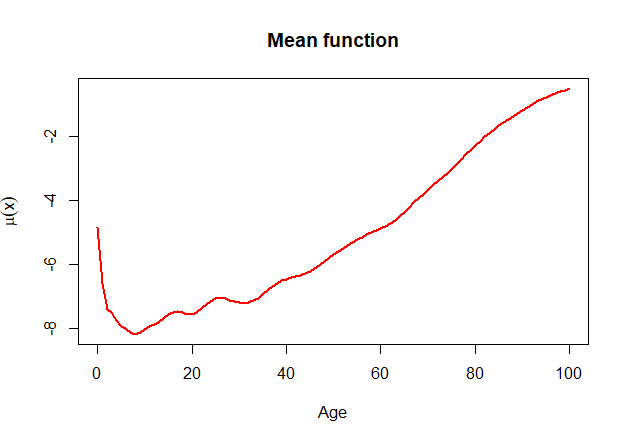}
		\includegraphics[width=0.4\linewidth]{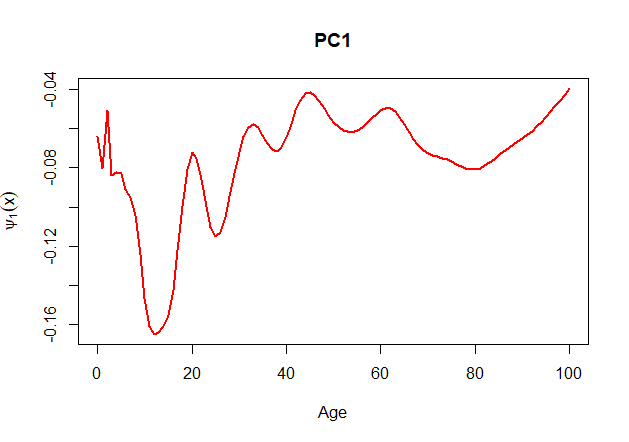}
	\end{minipage}
	\begin{minipage}{1\textwidth}
		\centering
		\includegraphics[width=0.4\linewidth]{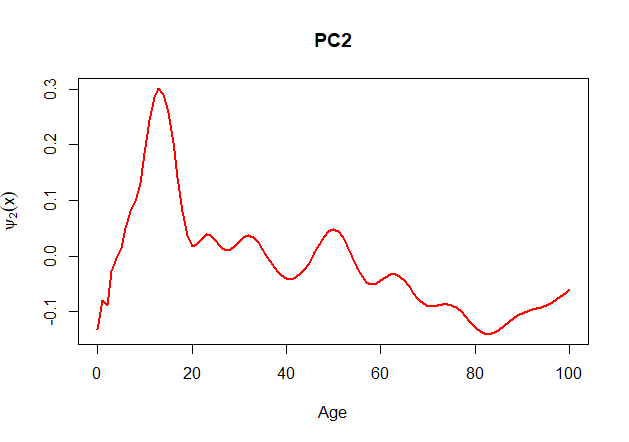}
		\includegraphics[width=0.4\linewidth]{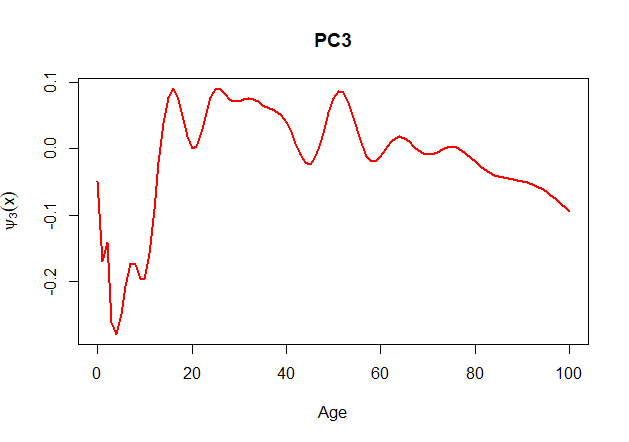}
		\subcaption{Female}
	\end{minipage}
	\caption{Estimated mean functions and the first three functional principal components for (a) male and (b) female mortality of Japan. }
	\label{fig: Estimated mean functions and the first three functional principal components for male and female}
\end{figure}

\begin{figure}[!thb]
	\ContinuedFloat
	\begin{minipage}{1\textwidth}
		\centering
		\includegraphics[width=0.4\linewidth]{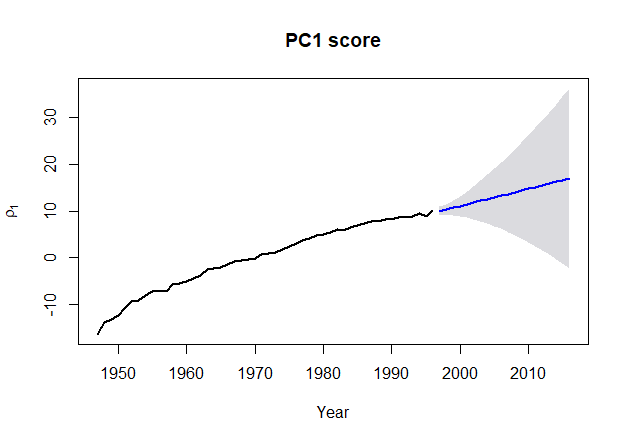}
		\includegraphics[width=0.4\linewidth]{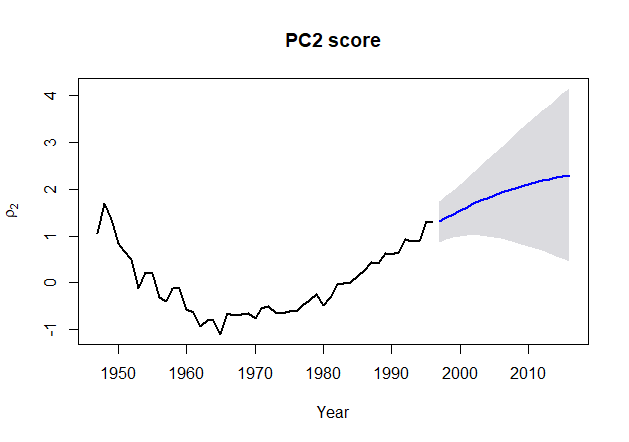}
	\end{minipage}
	\begin{subfigure}{\linewidth}
		\centering
		\includegraphics[width=0.4\linewidth]{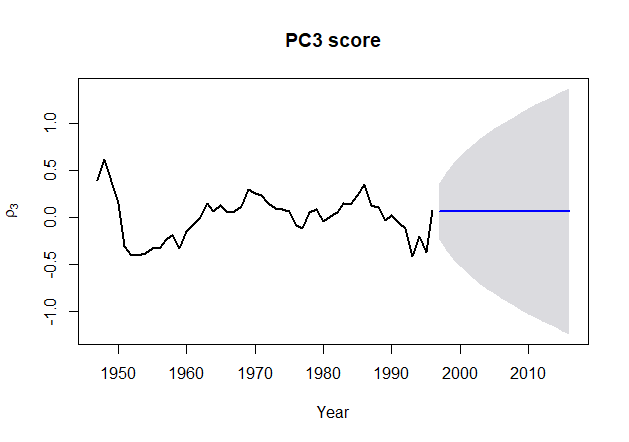}
		\subcaption{PC scores for male and female}
	\end{subfigure}
	\caption{(\textit{cont.}) (c) Corresponding estimated PC scores with 20-years-ahead forecast means with 95\% prediction intervals (in grey).}
\label{fig: Corresponding estimated scores of the PCs with a 20-years-ahead forecast}
\end{figure}

\begin{figure}[!thb]
	\begin{minipage}{1\textwidth}
		\centering
		\includegraphics[width=0.41\linewidth]{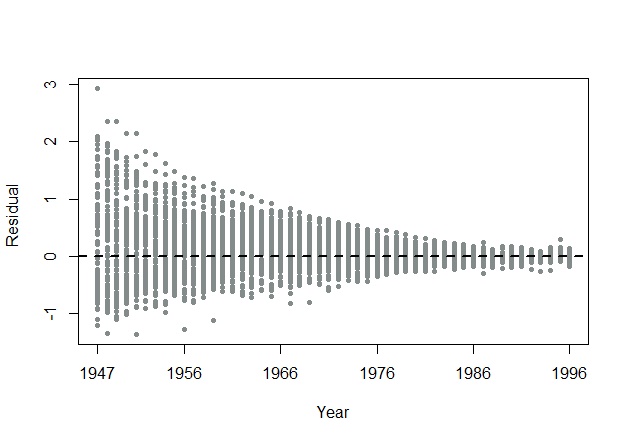}
		\includegraphics[width=0.41\linewidth]{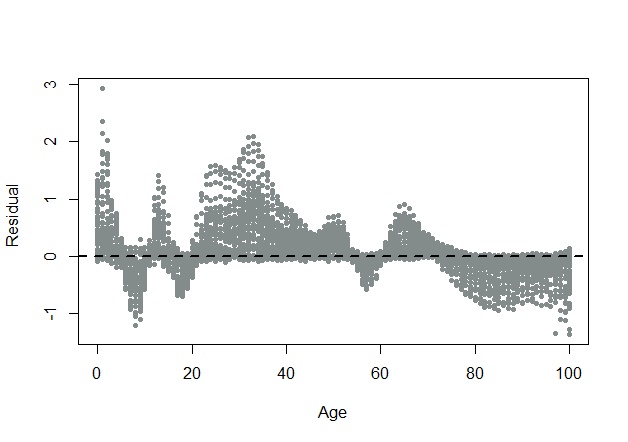}
		\subcaption{Male}
	\end{minipage}
	\begin{minipage}{1\textwidth}
		\centering
		\includegraphics[width=0.41\linewidth]{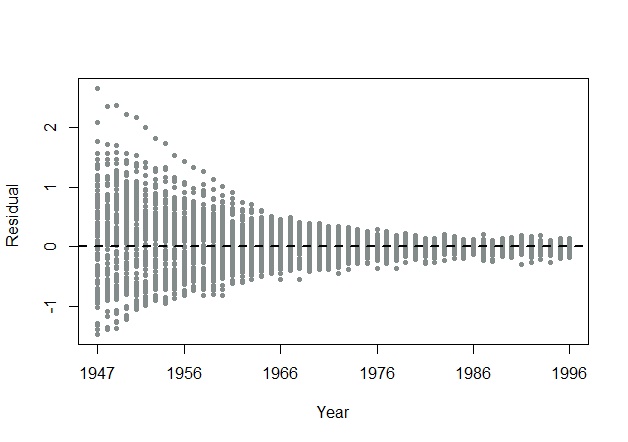}
		\includegraphics[width=0.41\linewidth]{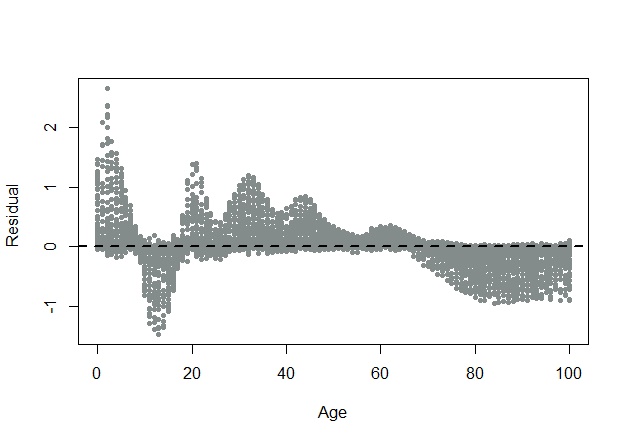}
		\subcaption{Female}
	\end{minipage}
	\caption{Residual plots by year and by age of the fitted sex-specific mortality rates of Japan for (a) male and (b) female via the wMFPCA model.}
	\label{fig: Residual plots by year and age of the fitted wMFPCA model for male and female mortality}
\end{figure}

\begin{figure}[!thb]
	\centering
	\begin{minipage}{0.7\textwidth}
		\centering
		\includegraphics[width=1\linewidth]{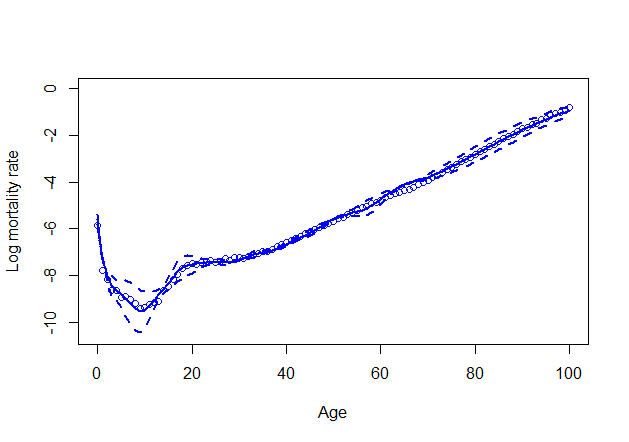}
		\subcaption{Male}
	\end{minipage}
	\begin{minipage}{0.7\textwidth}
		\centering
		\includegraphics[width=1\linewidth]{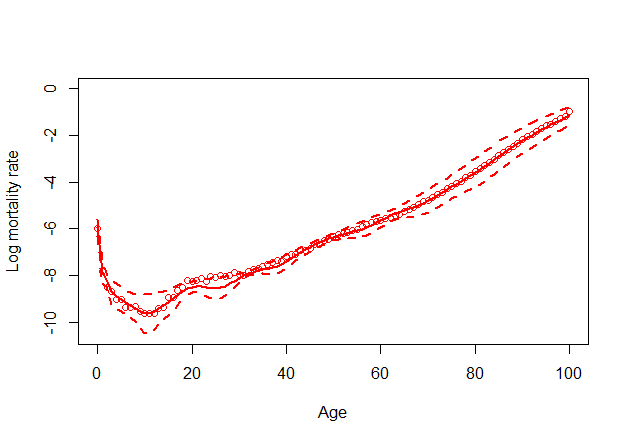}
		\subcaption{Female}
	\end{minipage}
	\caption{Fitted mortality curves for (a) male (with RMSE = 0.2006) and for (b) female (with RMSE = 0.2406) from age 0 to age 100 with 95\% confidence intervals using the wMFPCA model for the year 2016 based on the observations from the year 1947 to the year 1996 in Japan. Circles are the true log mortality rates, solid lines are the predictive means, and dashed lines are the 95\% confidence intervals.}
	\label{fig: Predicted mortality curves of male and female using the wMFPCA model}
\end{figure}
In the demonstration of the first proposed weighted MFPCA model, we aim to make  20-years-ahead out-of-sample forecasts for male and female mortality rates of Japan. We first split the dataset with the observed mortality rates from the year 1947 to the year 1996 and a test dataset with remaining observed mortality data from the year 1997 to the year 2016. We choose the weight parameter $\kappa$ in the wMFPCA model by minimising the average root mean square error (RMSE) stated in Equation (\ref{eq: Root mean square error}) based on rolling-window analyses. For example, in our demonstration, we aim to make a 20-years horizon prediction for the male and female mortality in Japan for the year 2016. We firstly use the data from the year 1947 to the year 1967 to make a 20-years horizon prediction for the mortality rates in the year 1987 then compute the RMSE with the observed data for the first rolling-window analysis. We subsequently use the data from the year 1947 to the year 1968 to make the prediction for the mortality rates in the year 1988 and compute RMSE with the observed data again. We repeat this procedure until it reaches a point where it uses the data from the year 1947 to the year 1976 to make a forecast for the endpoint of the observed data in the year 1996. We decide the value of the weight parameter over the interval $0 < \kappa < 1$ that can minimise the average RMSE of male and female mortality data among these ten rolling-window tests.
\par The mean functions for male and female and their functional principal components are estimated as discussed in the previous section. The analysis shows that the first three functional principal components for male and female explain 97.2\%, 2.3\% and 0.2\% respectively, and account for more than 99\% in total of the variations in the sample data. We, therefore, select the first three estimated principal components for approximations and demonstrations. For each score of the corresponding functional principal component shared by male and female, we forecast it independently by a univariate ARIMA time series using the R package `$\mathit{forecast}$' \citep{hyndman2007automatic}. The order of ARIMA models is chosen based on the Akaike information criterion (AIC).
\par The top panels of Figure \ref{fig: Estimated mean functions and the first three functional principal components for male and female} demonstrate the estimated mean functions $\hat{\mu}^{(i)}(x)$ and the first three estimated functional principal components $\{\hat{\psi}^{(i)}_{n}(x)\}$ for male $(i =$ M) and female $(i =$ F). Their corresponding scores of the PCs $\{\hat{\rho}_{t,n}\}^{3}_{n=1}$ with a 20-years-ahead out-of-sample forecast horizon and 95\% prediction intervals are displayed in the bottom panel of Figure \ref{fig: Corresponding estimated scores of the PCs with a 20-years-ahead forecast}. The functional principal components model different movements in mortality rates. The first functional principal components for male and female show a similar pattern, which both count for very high variation for young teenagers, then become level-off in middle-age and elderly. Although the corresponding scores for the first component show an increasing trend, we can interpret that there is a decreasing trend in mortality rates across our observed period given its corresponding first components for male and female are both negative. The second and the third principal components for male have relatively high negative variation among younger ages, but they get less oscillated in the ranges of middle-age and elderly. While the second and the third principal components for female show kinds of the opposite direction of each other.  
\par Figure \ref{fig: Residual plots by year and age of the fitted wMFPCA model for male and female mortality} presents the residual plots by year and by age of the wMFPCA model fitted by gender in Japan. As we have applied the weighting approach for placing more recent data into considerations, both the residual plots by year for male and female, have a funnel shape. These indicate that the fitted wMFPCA model captures more information from the recent data and leave the distant past data with less impact. The residual plots by age for male and female show that the fitted values of the models are underestimated for the young age groups and slightly overestimated for the elderly groups. These may be due to the higher mortality improvement rates for the young group and the lower mortality improvement rates for the elderly group over our fitting period. Figure \ref{fig: Predicted mortality curves of male and female using the wMFPCA model} shows the 20-years-ahead forecasting results of mortality curves of male (with RMSE = 0.2006) and female (with RMSE = 0.2406) from age 0 to age 100 for the year 2016 by the wMFPCA model based on the in-sample data from the year 1947 to the year 1996 in Japan.
\subsection{Sex-specific mortality forecasting by the coherent wMFPCA model}
The presentation of the coherent wMFPCA model forecasting follows the same strategies of how we split the dataset for in- and out-of-sample data and choosing the weight parameter for the coherent wMFPCA model as we have done for the wMFPCA model in the previous section. 
\par In the top panels of Figure \ref{fig: The estimated overall mean function and the first three estimated common trend functional principal components for total mortality}, we display the estimated overall mean function $\hat{\mu}(x)$ and the first three estimated functional principal components $\{\hat{\phi}_{k}(x)\}^{3}_{k=1}$. The corresponding scores of the first three estimated functional principal components $\{\hat{\beta}_{t,k}\}^{3}_{k=1}$ are presented along with 20-years-ahead out-of-sample forecast means and 95\% confidence intervals by ARIMA models in the bottom panel of Figure \ref{fig: Corresponding scores of the estimated common trend PCs with a 20-years-ahead forecast}. The first three common trend functional principal components capture 98\%, 1.7\% and 0.2\% variations respectively and more than 99\% of the variations in the age-specific total mortality overall among this dataset. Each functional principal component models different movements in mortality rates. $\hat{\phi}_{1}(x)$ primarily models the degree of variations in mortality among different age groups as we expect that the variance of mortality rates in young age groups is relatively larger than elderly groups. In contrast, $\hat{\phi}_{2}(x)$ and $\hat{\phi}_{3}(x)$ model the young adult age below 40 and differences between late teen and those over 60. From the first forecast common principal component scores, the declines of mortality rates seem to continue, and we do not spot any apparent pattern from the second and third estimated common principal component scores.
\par In the top panels of Figure \ref{fig: Estimated deviation functions from overall mean and the first three estimated deviation trend functional principal components for male and female}, we plot the estimated male and female deviation functions from the overall mean. It is obvious that the estimated male deviation function $\hat{\eta}^{\text{M}}(x)$ is a positive function, while the female estimated deviation function $\hat{\eta}^{\text{F}}(x)$ covers whole negative range over all ages. These indicate that the male mortality rates are in general higher than female's, and the difference reaches its peak at around age 20. All estimated male and female deviation functions from the overall mean behave like an upside-down reflection of each other and demonstrate completely reverse patterns. The bottom panels of Figure \ref{fig: Corresponding scores of the estimated deviation trend PCs with a 20-years-ahead forecast} show the first three estimated deviation trend functional principal components for male and female. From the forecasts of the three shared principal component scores among two genders, they seem to have a flat zero-convergent trend. It is therefore likely to achieve non-divergent forecasts between male and female subpopulations in the long term, in which we will further examine this in detail in the next section.
\par
\begin{figure}[!thb]
	\begin{minipage}{1\textwidth}
		\centering
		\includegraphics[width=0.41\linewidth]{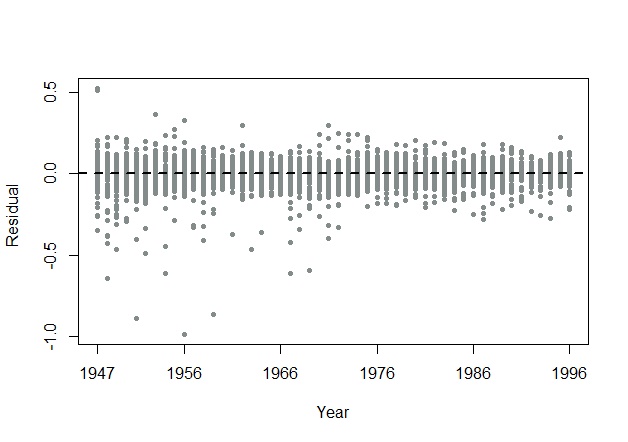}
		\includegraphics[width=0.41\linewidth]{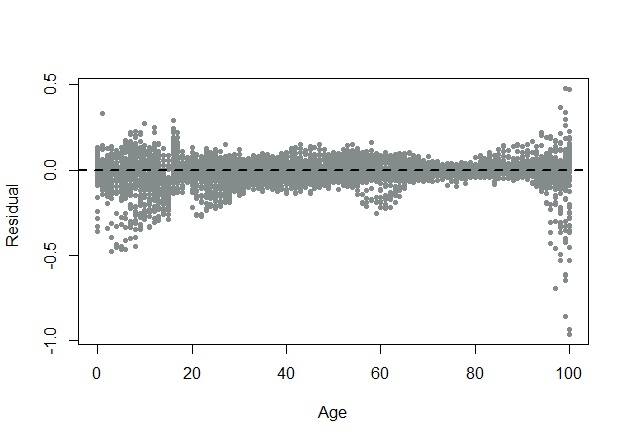}
		\subcaption{Male}
	\end{minipage}
	\begin{minipage}{1\textwidth}
		\centering
		\includegraphics[width=0.41\linewidth]{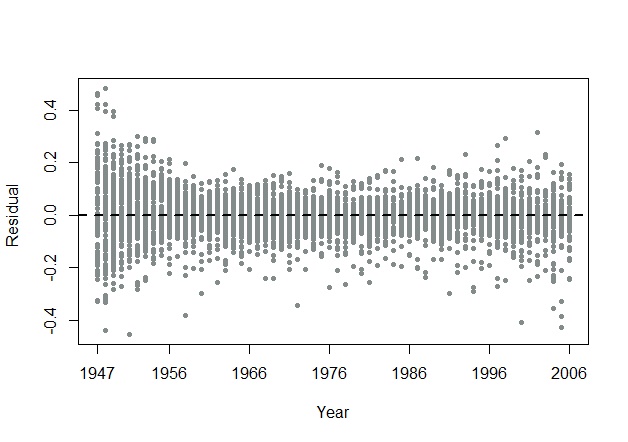}
		\includegraphics[width=0.41\linewidth]{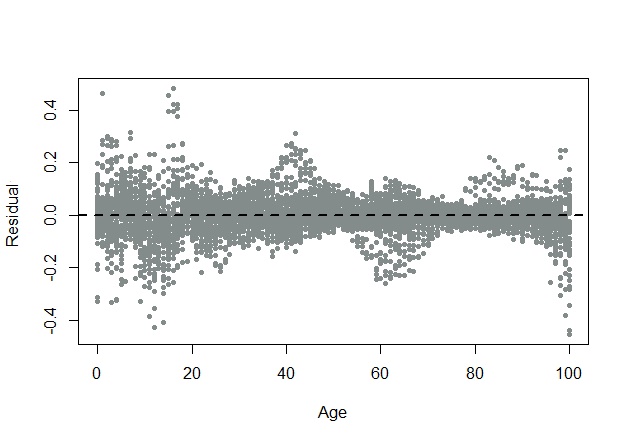}
		\subcaption{Female}
	\end{minipage}
	\caption{Residual plots by year and by age of the fitted sex-specific mortality rates of Japan for (a) male and (b) female via the coherent wMFPCA model.}
	\label{fig: Residual plots by year and age of the fitted coherent wMFPCA model for male and female mortality}
\end{figure}

\begin{figure}[!thb]
	\centering
	\begin{minipage}{0.7\textwidth}
		\centering
		\includegraphics[width=1\linewidth]{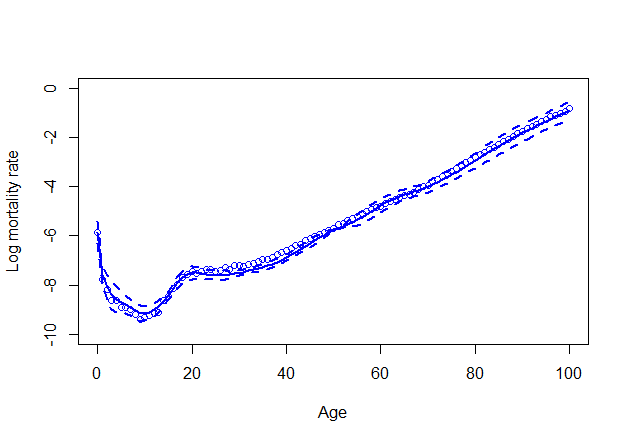}
		\subcaption{Male}
	\end{minipage}
	\begin{minipage}{0.7\textwidth}
		\centering
		\includegraphics[width=1\linewidth]{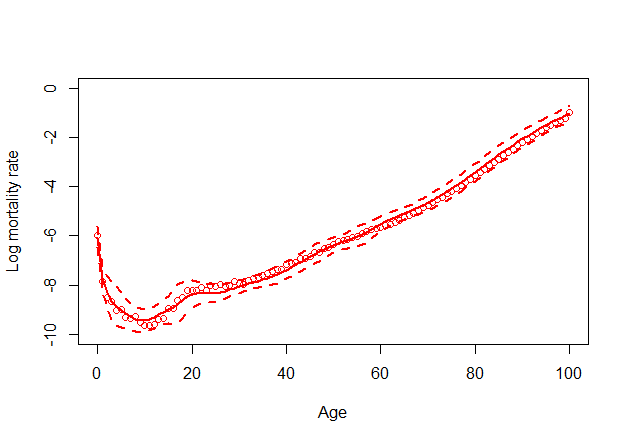}
		\subcaption{Female}
	\end{minipage}
	\caption{Predicted mortality curves of male (with RMSE = 0.1821) and female (with RMSE = 0.1460) from age 0 to age 100 with 95\% confidence intervals using the coherent wMFPCA model for the year 2016 based on the observations from the year 1947 to the year 1996 in Japan. Circles are the true log mortality rates, solid lines are the predictive means and dashed lines are the 95\% confidence intervals.}
	\label{fig: Predicted mortality curves of male and female using the coherent wMFPCA model}
\end{figure}

 Figure \ref{fig: Residual plots by year and age of the fitted coherent wMFPCA model for male and female mortality} shows the residual plots by year and by age of the fitted coherent wMFPCA model for male and female mortality of Japan. Compared to the residual patterns of the fitted wMFPCA model used in the previous section, the fitted coherent wMFPCA model seems to capture mortality information evenly over the whole observed period, given that both residual plots demonstrate i.i.d. random variables patterns with zero mean and equal variance across all the age-groups over our examined period. Figure \ref{fig: Predicted mortality curves of male and female using the coherent wMFPCA model} shows the 20-years-ahead forecasting results of mortality curves of male (with RMSE = 0.1821) and female (with RMSE = 0.1460) from age 0 to age 100 for the year 2016 using the coherent wMFPCA model based on the in-sample data from the year 1947 to the year 1996 in Japan.
\begin{figure}[!thb]
	\begin{minipage}{1\textwidth}
		\centering
		\includegraphics[width=0.4\linewidth]{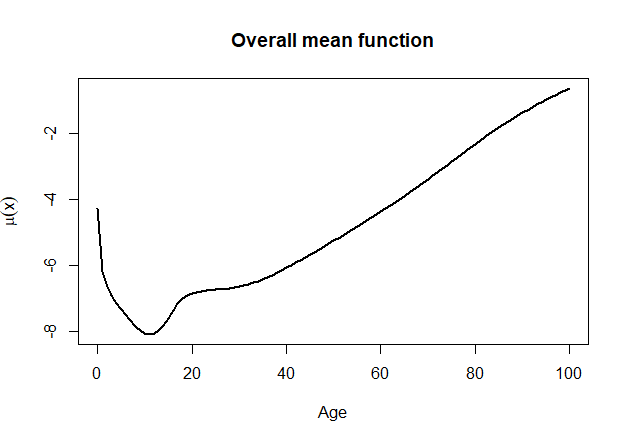}
		\includegraphics[width=0.4\linewidth]{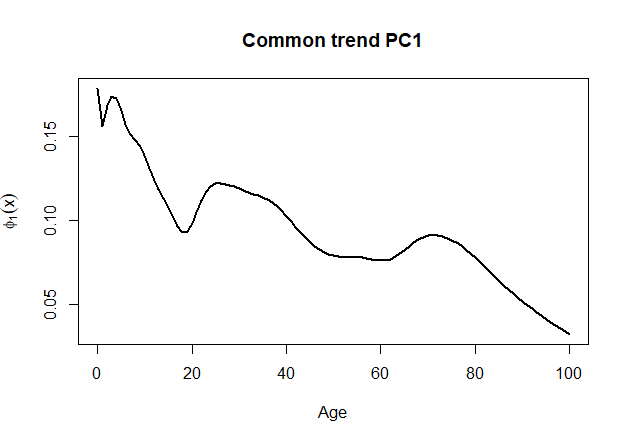}
	\end{minipage}
	\begin{minipage}{1\textwidth}
		\centering
		\includegraphics[width=0.4\linewidth]{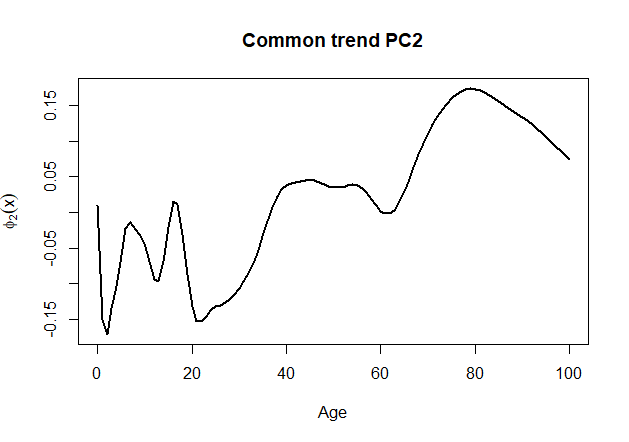}
		\includegraphics[width=0.4\linewidth]{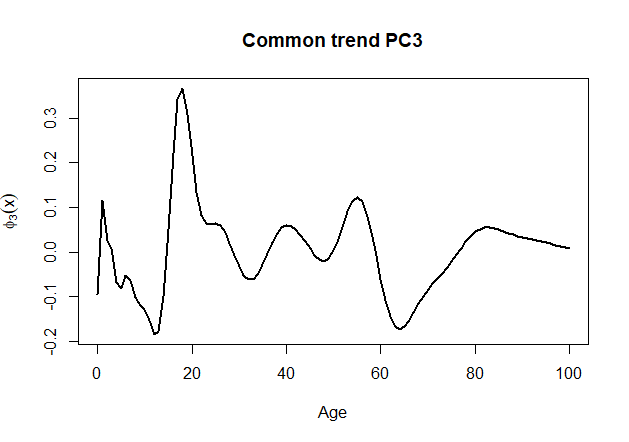}
	\end{minipage}
	\caption{Estimated overall mean function and the first three estimated common trend functional principal components for the total mortality rates of Japan.}
	\label{fig: The estimated overall mean function and the first three estimated common trend functional principal components for total mortality}
\end{figure}
\begin{figure}[!thb]
	\ContinuedFloat
	\begin{minipage}{1\textwidth}
		\centering
		\includegraphics[width=0.4\linewidth]{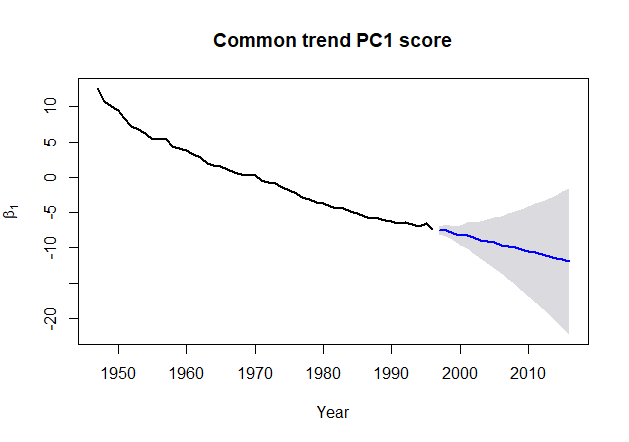}
		\includegraphics[width=0.4\linewidth]{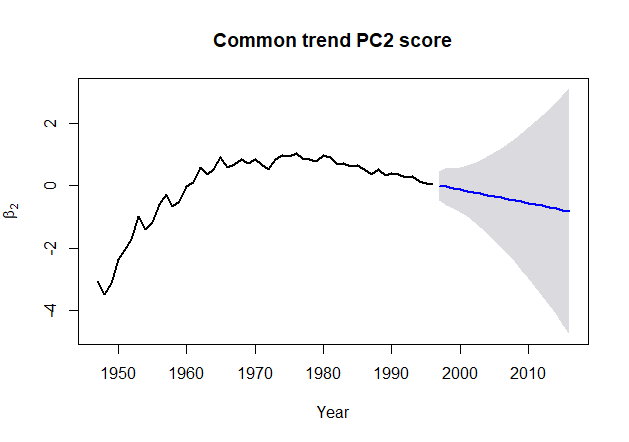}
	\end{minipage}
	\begin{subfigure}{\linewidth}
		\centering
		\includegraphics[width=0.4\linewidth]{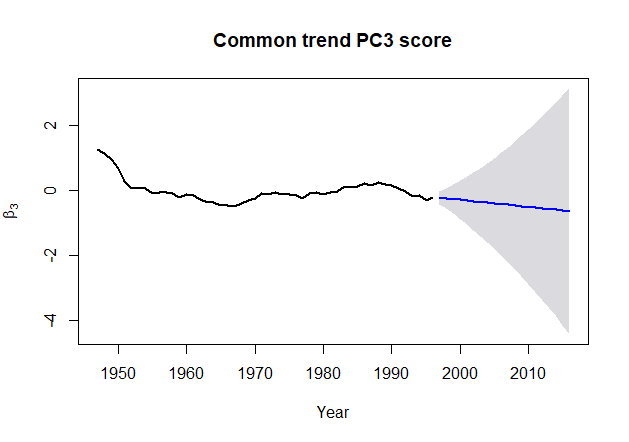}
	\end{subfigure}
	\caption{(\textit{cont.}) Corresponding PC scores of the estimated common trend with 20-years-ahead forecast means with 95\% confidence intervals (in grey).}
	\label{fig: Corresponding scores of the estimated common trend PCs with a 20-years-ahead forecast}
\end{figure}
\begin{figure}[!thb]
	\begin{minipage}{1\textwidth}
		\centering
		\includegraphics[width=0.4\linewidth]{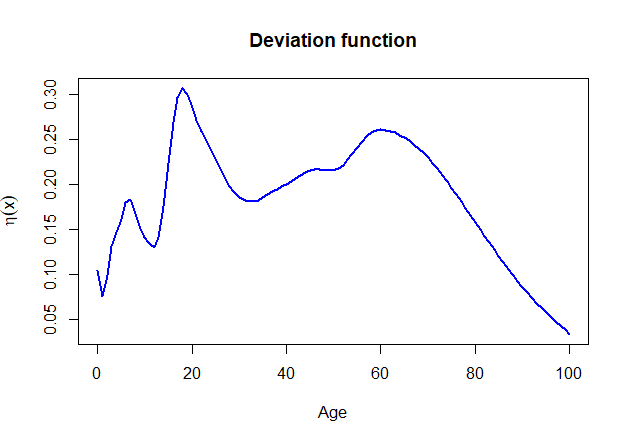}
		\includegraphics[width=0.4\linewidth]{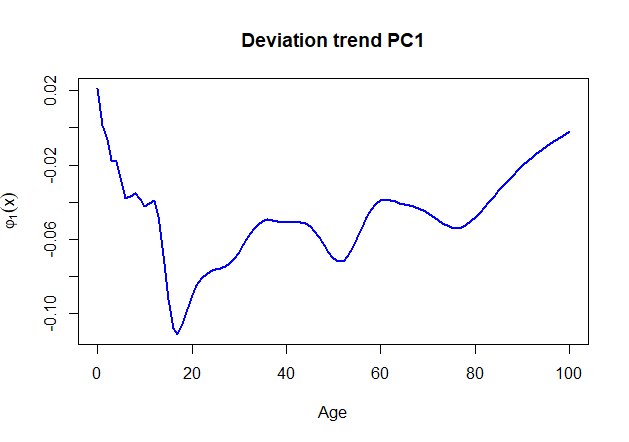}
	\end{minipage}
	\begin{minipage}{1\textwidth}
		\centering
		\includegraphics[width=0.4\linewidth]{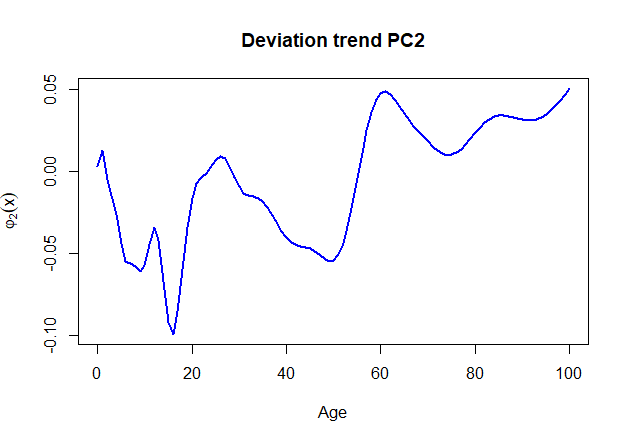}
		\includegraphics[width=0.4\linewidth]{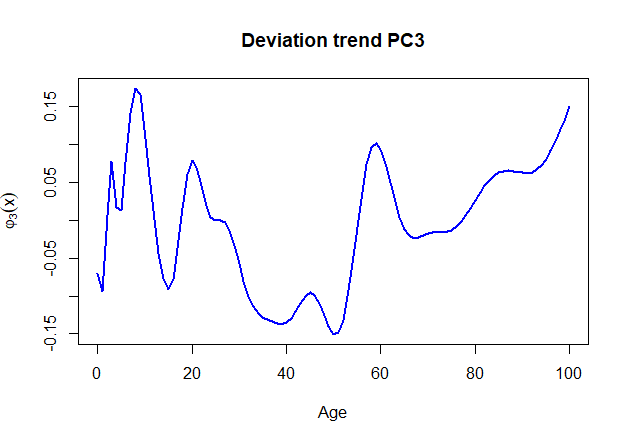}
		\subcaption{Male deviation from the overall mean function}
	\end{minipage}
	\begin{minipage}{1\textwidth}
		\centering
		\includegraphics[width=0.4\linewidth]{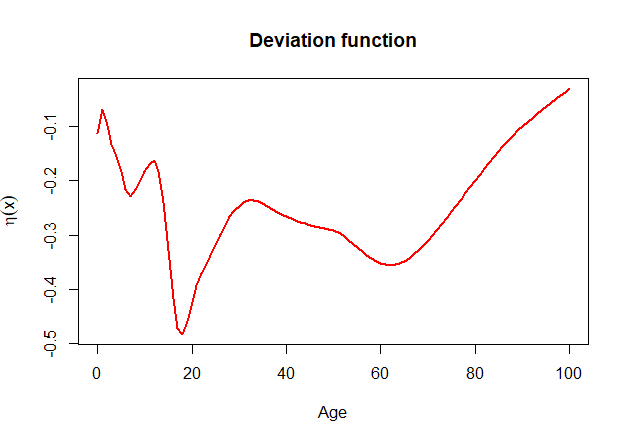}
		\includegraphics[width=0.4\linewidth]{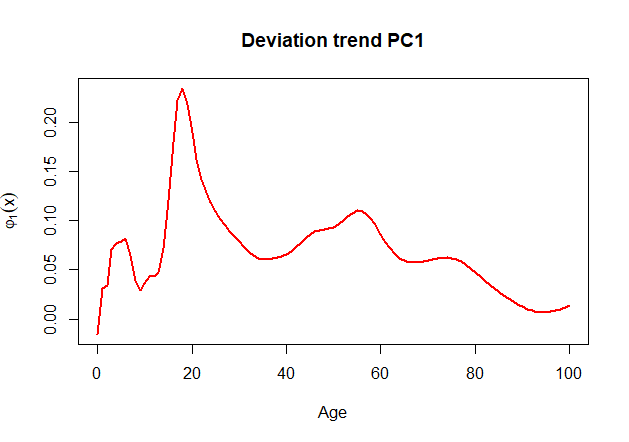}
	\end{minipage}
	\begin{minipage}{1\textwidth}
		\centering
		\includegraphics[width=0.4\linewidth]{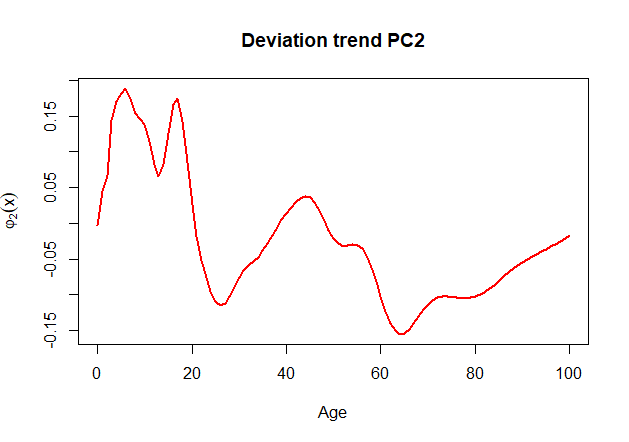}
		\includegraphics[width=0.4\linewidth]{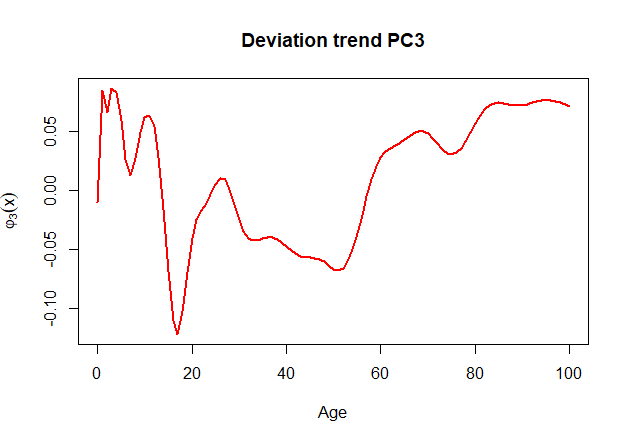}
		\subcaption{Female deviation from the overall mean function}
	\end{minipage}
	\caption{Estimated deviation functions from the overall mean function and the first three estimated deviation trend functional principal components for (a) male and (b) female mortality rates of Japan.}
	\label{fig: Estimated deviation functions from overall mean and the first three estimated deviation trend functional principal components for male and female}
\end{figure}
\begin{figure}[!thb]
	\ContinuedFloat
	\begin{minipage}{1\textwidth}
		\centering
		\includegraphics[width=0.4\linewidth]{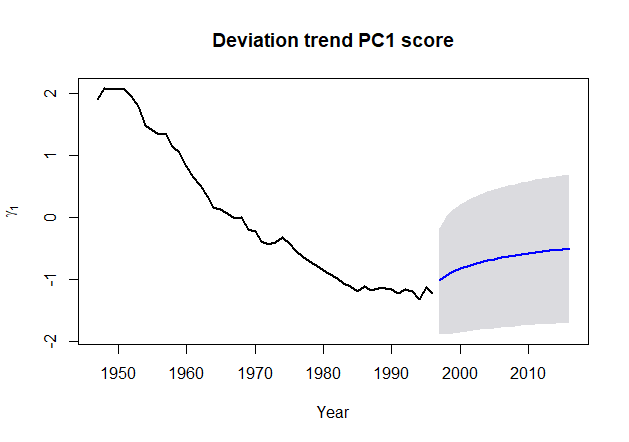}
		\includegraphics[width=0.4\linewidth]{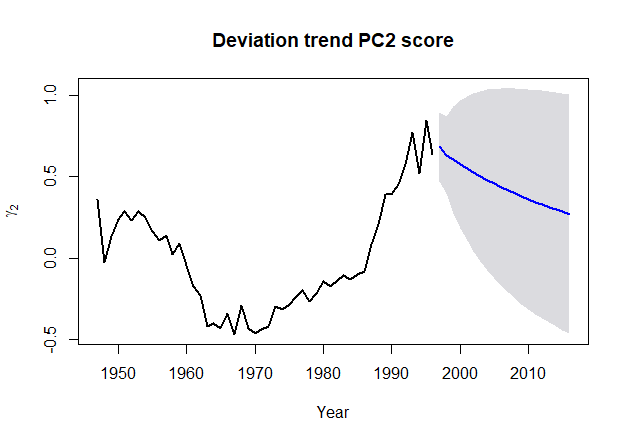}
	\end{minipage}
	\begin{subfigure}{\linewidth}
		\centering
		\includegraphics[width=0.4\linewidth]{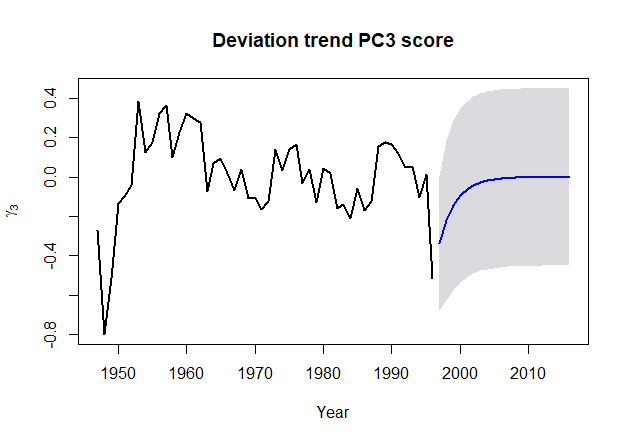}
	\end{subfigure}
	\caption{(\textit{cont.}) (c) Corresponding PC scores of the estimated deviation trend with  20-years-ahead forecast means with 95\% confidence intervals (in grey).}
\label{fig: Corresponding scores of the estimated deviation trend PCs with a 20-years-ahead forecast}
\end{figure}

\subsection{Forecast pattern with mortality sex ratio and life expectancy by coherent and non-coherent forecasting methods}
In this section, we move to examine and compare forecast patterns with mortality sex ratios and life expectancy by the forecasts of the two proposed models $-$ the wMFPCA model and the coherent wMFPCA model with two existing models $-$ the independent FPCA model \citep{hyndman2009forecasting} and the Product-Ratio model \citep{hyndman2013coherent}. \par The independent FPCA model is a univariate FPCA method for forecasting two subpopulation groups independently without considerations of any potential correlation of two subpopulation groups. Similar to the wMFPCA model, the forecast results of the independent FPCA model are based on a non-stationary time series model on its estimated principal component scores, leading to forecast results of two or more subpopulations divergent to different directions in the long run. It is thus regarded as a non-coherent forecasting approach. Meanwhile, the Product-Ratio model begins with an idea of obtaining the product and ratio function of all subpopulations by assuming all subpopulations have equal variance. In the log scale, the product function can be treated as the sum of all sub populations, whereas the ratio function can be treated as the differences among subpopulations. The predictions can be obtained by firstly applying the independent FPCA model to forecast the future realisations of the product and ratio functions separately, then transforming the forecasts of the product and ratio functions back to the original subpopulations functions. The convergent forecasts are achieved by using stationary time series methods, namely the ARMA model or the ARFIMA model, on the ratio function, which implicitly implies that the differences among subpopulations will be convergent to zero in the long term. It is therefore viewed as an example of coherent forecasting approach. In a similar vein, the proposed coherent wMFPCA model restricts the stationary properties on the deviation functions of each subpopulation from the overall mean to accomplish the coherent forecasting with no need to assume all the subpopulations have the same variances.
\par
\begin{figure}[!thb]
	\begin{subfigure}{\linewidth}
	\centering
		\includegraphics[width=0.48\linewidth]{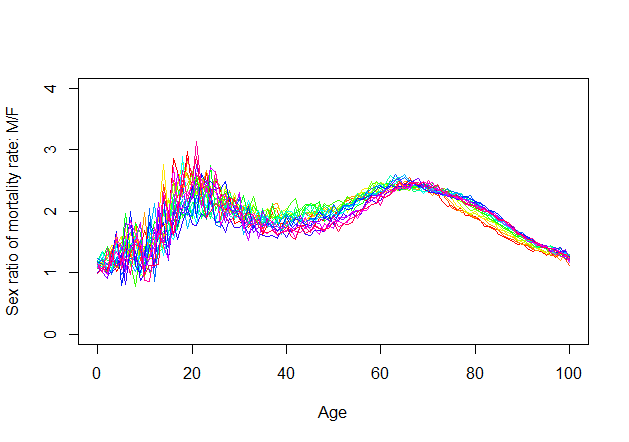}
		\subcaption{Historical data}
	\end{subfigure}
\par
	\begin{minipage}{0.5\textwidth}
		\centering
		\includegraphics[width=0.9\linewidth]{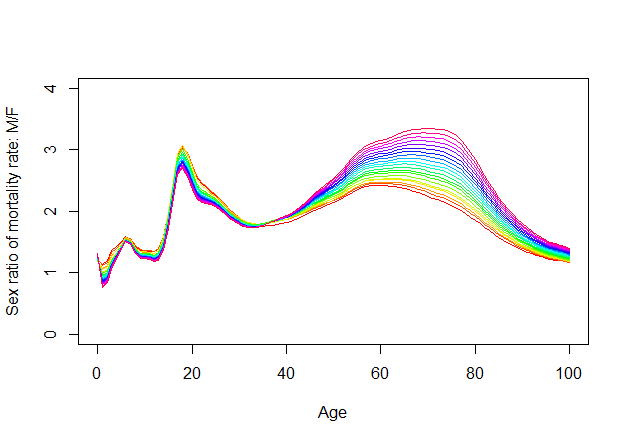}
		\subcaption{Independent FPCA model}
	\end{minipage}
	\begin{minipage}{0.5\textwidth}
	\centering
	\includegraphics[width=0.9\linewidth]{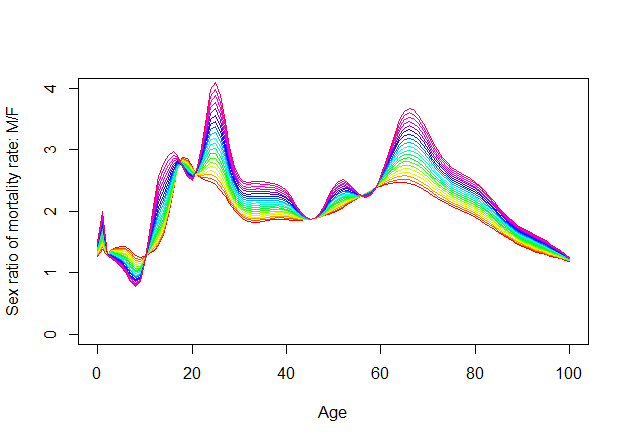}
	\subcaption{wMFPCA model}
\end{minipage}
\begin{minipage}{0.5\textwidth}
	\centering
	\includegraphics[width=0.9\linewidth]{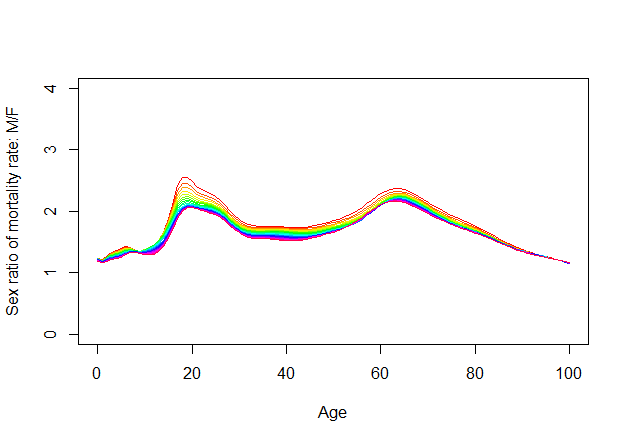}
	\subcaption{Product-Ratio model}
\end{minipage}
\begin{minipage}{0.5\textwidth}
	\centering
	\includegraphics[width=0.9\linewidth]{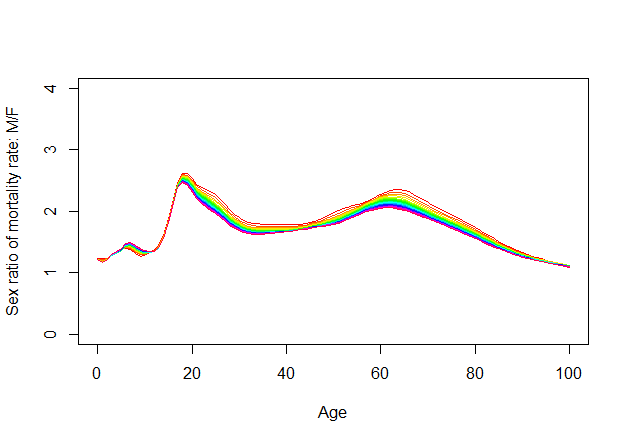}
	\subcaption{Coherent wMFPCA model}
\end{minipage}
	\caption{Historical Japanese mortality sex ratio and 20-years-ahead forecasts of mortality sex ratios in Japan from the year 1997 to the year 2016 using the independent FPCA model, the wMFPCA model, the Product-Ratio model and the coherent wMFPCA model.}
	\label{fig: Historical Japanese mortality sex ratio and 20-years-ahead forecasts of mortality sex ratios}
\end{figure}

\begin{figure}[!thb]
	\begin{minipage}{0.5\textwidth}
	\centering
	\includegraphics[width=0.9\linewidth]{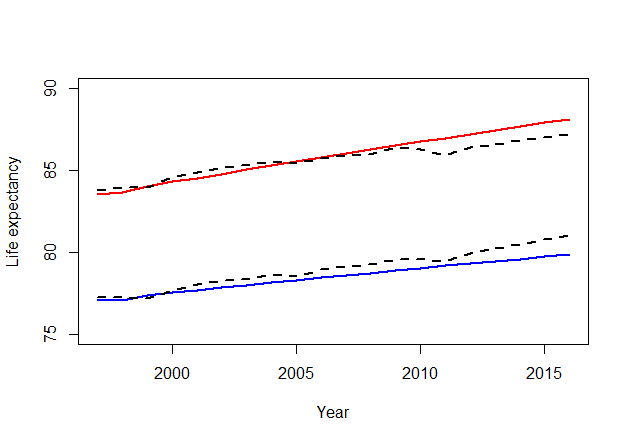}
	\subcaption{Independent FPCA model}
\end{minipage}
\begin{minipage}{0.5\textwidth}
	\centering
	\includegraphics[width=0.9\linewidth]{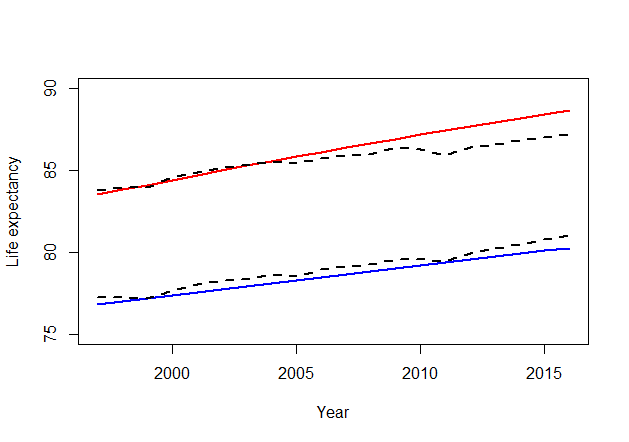}
	\subcaption{wMFPCA model}
\end{minipage}
\begin{minipage}{0.5\textwidth}
	\centering
	\includegraphics[width=0.9\linewidth]{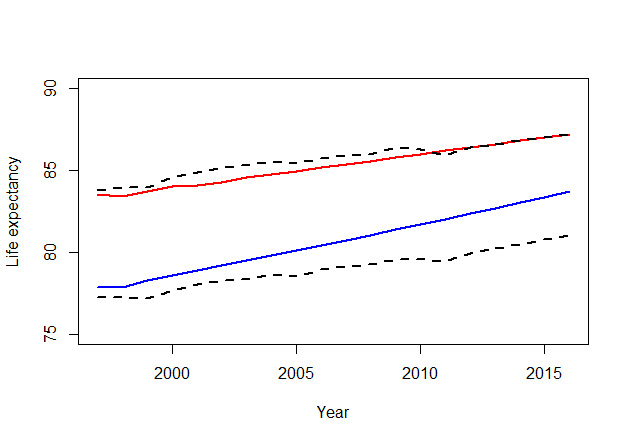}
	\subcaption{Product-Ratio model}
\end{minipage}
\begin{minipage}{0.5\textwidth}
	\centering
	\includegraphics[width=0.9\linewidth]{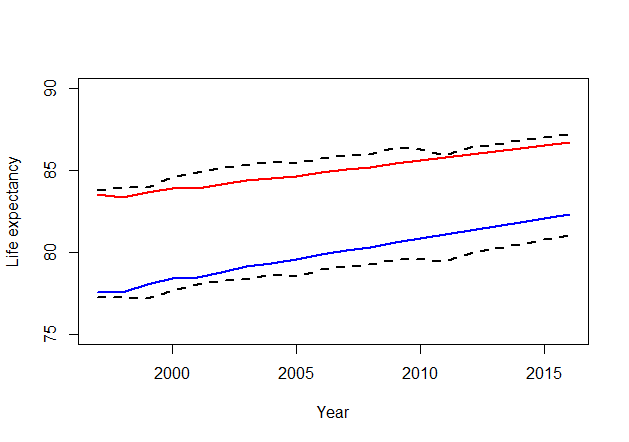}
	\subcaption{Coherent wMFPCA model}
\end{minipage}
\caption{20-year life expectancy predicted curves for male and female in Japan using the independent FPCA model, the wMFPCA model, the Product-Ratio model and the coherent wMFPCA model. Blue sold line is used for male and red sold line is used for female. Dotted lines are the observed life expectancy for males and females.}
\label{fig: 20-year life expectancy predicted curve for males and females}
\end{figure}

 To deliver the concept of coherent forecasting more concretely, we plot the historical mortality sex ratios (Male/Female) and the life expectancy curves obtained from the observed male and female mortality rates from the year 1997 to the year 2016 alongside the 20-years-ahead forecasts of the mortality sex ratios and the life expectancy curves from the year 1997 to the year 2016 by the non-coherent forecast methods $-$ the independent FPCA model and the wMFPCA model and the coherent forecast methods $-$ the Product-Ratio model and the coherent wMFPCA model using the observed mortality rates from the year 1947 to the year 1996 in Figure \ref{fig: Historical Japanese mortality sex ratio and 20-years-ahead forecasts of mortality sex ratios} and Figure \ref{fig: 20-year life expectancy predicted curve for males and females}. We can see that all the coherent forecasting methods exhibit a fairly one-to-one smooth pattern with the actual mortality sex ratio, and with much less fluctuation than the non-coherent forecast methods in Figure \ref{fig: Historical Japanese mortality sex ratio and 20-years-ahead forecasts of mortality sex ratios}. The convergent forecastings by the coherent models also fit well with the actual biological characteristics trends in Figure \ref{fig: 20-year life expectancy predicted curve for males and females}, where the differences in males and females life expectancy converge to a certain level gradually and slowly, instead of diverging into different directions like the forecast results of the non-coherent forecast methods. Our demonstrations prove the importance of coherent modelling when there exist common biological characteristics among several subpopulations. 
      
\subsection{Forecast accuracy evaluation with comparisons to other existing methods}
We now evaluate and compare the forecast accuracy of our two proposed models $-$ the wMPFCA model and the coherent wMFPCA model with the two existing models $-$  the independent FPCA model and the Product-Ratio model demonstrated in the previous section. In order to have a comprehensive investigation of the forecast accuracy of our two proposed models, we consider ten other developed countries for which data are also available in the Human Mortality Database. We restrict data periods of all selected countries commencing in the year 1947 up to the year 2016 (70 years in total) for a unified purpose. We examine and quantify the forecasting performance of our models by a rolling window analysis, which is frequently used for assessing the consistency of a model's forecasting ability by rolling a fixed size prediction interval (window) throughout the observed sample \citep{zivot2007modeling}. More generally speaking, we hold the sample data from the initial year up to the year $t$ as holdout samples. We produce the forecast for the $t + h$ year where $h$ is the forecast horizon, then determine the forecasts errors by comparing the forecast result with the actual out-of-sample data. We increase one rolling window (1 year ahead) in year $t + 1$ to make the same procedure again for the year $t + h + 1$ until the rolling window analysis covers all available data. 
\par We include four different forecast horizons $(h = 5, 10, 15$ and $20)$ with ten sets of rolling-window to exam the short-term, the mid-term and the long-term forecast abilities of the proposed two models. We use the root mean square error (RMSE) to measure the standard deviation of the average square prediction error regardless of sign. In our experiments, we define the RMSE as follows:
\[
\text{RMSE}^{(i)}_{c}(h) = \sqrt{\frac{1}{10\times 101}\sum_{w=0}^{9}\sum_{j=1}^{101}\bigg(Y^{(i)}_{t+w+h}(x_{j}) - \hat{Y}^{(i)}_{t+w+h}(x_{j}) \bigg)^{2}},
\]   
where $c$ is the selected country, $w$ is the rolling-window set, $i$ is the subpopulation for male $(i =$ M) and for female $(i =$ F) and $j$ is the age group including from age 0 to age 100 in our experiment. 
\par Based on the average RMSE results of ten sets of rolling window analysis across ten countries in four different forecast horizons presented in Table \ref{table: Forecast accuracy of mortality for male and female by the average RMSEs in ten sets of rolling-windows analysis}, the forecast performances of the wMFPCA model and the Product-Ratio model are reasonably similar with no particular outstanding area. The independent FPCA model performs consistently the most desirable for forecasting female mortality. Small variabilities of female mortality across all observed countries over age and time may mainly contribute to the superiority of the independent FPCA model over the others in our experiments. Meanwhile, the proposed coherent wMFPCA method performs the best in terms of having the least male forecast errors. The coherent wMFPCA method seems to be capable of capturing rapid changes in male mortality across different periods and age groups in many tested countries.  
\par When we consider the forecast horizon size up to fifteen or twenty years, we can see that the coherent models maintain relatively less forecast bias among two sexes than the non-coherent models. For instance, the independent FPCA model produces comparatively large forecasting errors for male mortality and gives small forecast inaccuracies for female death rates. In contrast, the coherent models can keep the same level of forecast errors for both. With assumed joint biological characteristics among the two genders that we discussed in the previous section, the mortality pattern among two sexes is supposed to get similar in the long run, and the convergent designed forecast model is therefore needed. In particular, the coherent wMFPCA model is proved to be more suitable and accurate than the Product-Ratio model as it produces the smallest overall forecast errors and bias for both genders in the long-term prediction in our study. 
\par The main finding in this section is that in the two-sex case, the accuracy of the male forecast is significantly improved by the coherent wMFPCA model at the small expense of accuracy in female mortality forecasts. By adopting the coherent wMFPCA model, the forecast accuracy among all subpopulations is homogeneous as it incorporates additional information into the forecast for a single subpopulation. The additional information acts as a frame of reference limiting to the probability of a subpopulation forecast which may continue a diverging trend from other related subpopulations directions. This feature of the coherent wMFPCA model is also useful in some other specific practical applications, such as financial planning with several related stock prices, in a situation that we aim to maintain a balanced error margin amongst all subpopulations. This speciality is unique and has not been achieved by other non-coherent or single population models.      
\begin{landscape}
\setlength{\belowcaptionskip}{1pt} 
\begin{table}[ht]
\begin{tabular}{cccccccccccccccc}  
\toprule
\multirow{2}{*}{
\parbox[c]{.06\linewidth}{\centering Country}}
& \multicolumn{3}{c}{Independent FPCA model} &&
\multicolumn{3}{c}{wMFPCA model} &&
\multicolumn{3}{c}{Product-Ratio model} && 
\multicolumn{3}{c}{Coherent wMFPCA model} \\ 
\cmidrule{2-4} \cmidrule{6-8} \cmidrule{10-12} \cmidrule{14-16}
& {\centering M} & {F} & {$\frac{\text{M+F}}{2}$} && {M} & {F} & {$\frac{\text{M+F}}{2}$} && {M} & {F} & {$\frac{\text{M+F}}{2}$} && {M} & {F} & {$\frac{\text{M+F}}{2}$}\\
\midrule
\underline{$h = 5$}\\
Australia & 0.1309& \cellcolor{yellow} 0.1771   & 0.1540   && 0.1305  & 0.1775    & \cellcolor{yellow} 0.1540   && \cellcolor{yellow}0.1303 & 0.1782 & 0.1543 && 0.1384 & 0.1797  & 0.1591 \\
Belgium& 0.2183    & \cellcolor{yellow}0.2499   & \cellcolor{yellow}0.2341    && 0.2365    & 0.2661    & 0.2513 && 0.2583 & 0.2820 & 0.2702 && \cellcolor{yellow}0.2150 & 0.2561   & 0.2355 \\
Canada&0.1328 & 0.1428   & 0.1378    && 0.1328     & \cellcolor{yellow} 0.1342 & 0.1335 && 0.1330   & 0.1387 & 0.1359&& \cellcolor{yellow} 0.1250 & 0.1379  & \cellcolor{yellow} 0.1315 \\
France  & 0.1101    & 0.1196   & 0.1149    && 0.1172 & 0.1197    & 0.1184    && \cellcolor{yellow} 0.1057   & \cellcolor{yellow} 0.1109 & \cellcolor{yellow} 0.1083 && 0.1090    & 0.1144  & 0.1117 \\
Italy & 0.1371    & \cellcolor{yellow}0.1331   & 0.1351    && 0.1575     & 0.1427    & 0.1501    && 0.1425 & 0.1456 & 0.1440 && \cellcolor{yellow} 0.1230 & 0.1397& \cellcolor{yellow} 0.1313\\
Japan & 0.1305    & 0.1439   & 0.1372    && \cellcolor{yellow} 0.1103 & \cellcolor{yellow} 0.1227 & \cellcolor{yellow} 0.1165&& 0.1276    & 0.1420 & 0.1348 && 0.1248    & 0.1444    & 0.1346 \\
Netherlands& 0.2059 & \cellcolor{yellow} 0.2025& 0.2042 && 0.2153& 0.2138& 0.2145&& \cellcolor{yellow} 0.2016& 0.2079 & 0.2048 && 0.2025& 0.2051 &\cellcolor{yellow} 0.2038 \\
Spain & \cellcolor{yellow} 0.1652& \cellcolor{yellow}0.1486&\cellcolor{yellow}0.1569  && 0.2200   & 0.1805& 0.2003  && 0.1874& 0.1684 & 0.1779&& 0.1668 & 0.1561 & 0.1615\\
U.K  & 0.1321 & 0.1176& 0.1249&& 0.1966 & 0.1473& 0.1720&& \cellcolor{yellow} 0.1274 &\cellcolor{yellow}0.1141 & \cellcolor{yellow} 0.1207&& 0.1298 & 0.1201 & 0.1250\\
U.S.A & \cellcolor{yellow} 0.0829& 0.0844   & 0.0836&& 0.0923 & 0.0890& 0.0907    && 0.0874& \cellcolor{yellow} 0.0774
& \cellcolor{yellow} 0.0824&& 0.0943 & 0.0787 & 0.0865\\
Average & 0.1446 & \cellcolor{yellow} \textbf{0.1520}& 0.1483&& 0.1609& 0.1593
& 0.1601&& 0.1501 & 0.1565& 0.1533&& \cellcolor{yellow} \textbf{0.1429}& 0.1532   & \cellcolor{yellow} \textbf{0.1480}\\
\midrule
\underline{$h = 10$}\\
Australia&0.2053&\cellcolor{yellow}0.2026& \cellcolor{yellow}0.2040 && 0.2086&0.2265& 0.2175&&\cellcolor{yellow}0.1976& 0.2109&0.2043&&0.2164&0.2248&0.2206\\
Belgium&\cellcolor{yellow}0.2598&\cellcolor{yellow}0.2560&\cellcolor{yellow}0.2579&&0.3066& 0.3034&0.3050
&& 0.3174 & 0.2986 & 0.3080&& 0.2739 & 0.2979 & 0.2859\\
Canada&0.1663 & 0.1715 & 0.1689 && 0.1841 & 0.1708 & 0.1774&&\cellcolor{yellow}0.1560 &\cellcolor{yellow}0.1671
& \cellcolor{yellow}0.1616&& 0.1614 & 0.1715 & 0.1665\\
France&0.1638& 0.1675& 0.1656 && 0.2048& 0.1789 & 0.1918&&\cellcolor{yellow}0.1601
& \cellcolor{yellow} 0.1502&\cellcolor{yellow} 0.1551 && 0.1780& 0.1778 &0.1779\\
Italy& 0.2513 & \cellcolor{yellow}0.1558&\cellcolor{yellow}0.2036 && 0.3185 & 0.2522 & 0.2854  && 0.2765 & 0.2656& 0.2710 &&\cellcolor{yellow}0.2371&0.2402&0.2387 \\
Japan& 0.2217 & 0.1803& 0.2010 && \cellcolor{yellow}0.1605&\cellcolor{yellow}0.1565& \cellcolor{yellow}0.1585&&0.1941 & 0.1763&0.1852&&0.1638&0.1731&0.1684 \\
Netherlands  & 0.2659& \cellcolor{yellow}0.2317&0.2488&&0.3184&0.2984&0.3084 &&\cellcolor{yellow}0.2529 & 0.2384
& \cellcolor{yellow}0.2457&& 0.2605 & 0.2407 & 0.2506\\
Spain  & 0.3134 & 0.2059 & \cellcolor{yellow}0.2597&&0.4188&0.3044&0.3616 && 0.3393 & 0.3093 & 0.3243 && 0.2932 & 0.2680 & 0.2806 \\
U.K&0.1816&0.1410&0.1613&&0.1605&0.1565& 0.1585&&\cellcolor{yellow}0.1681 &\cellcolor{yellow}0.1298
& \cellcolor{yellow}0.1489 && 0.1595& 0.1459 &0.1527\\
U.S.A & \cellcolor{yellow}0.1290 & 0.1423& 0.1357&& 0.1592& 0.1508&0.1550&& 0.1374& \cellcolor{yellow}0.1207& 0.1291
&& 0.1361& 0.1246& 0.1304 \\
Average& 0.2158 & \cellcolor{yellow}\textbf{0.1855} & \cellcolor{yellow}\textbf{0.2006}  && 0.2440  & 0.2198 & 0.2319 &&0.2199& 0.2067 & 0.2133 && \cellcolor{yellow}\textbf{0.2080}   & 0.2064   & 0.2072 \\
\bottomrule
\end{tabular}
\caption{Forecast accuracy of mortality for male and female using the independent FPCA model, the wMFPCA model, the Product-Ratio model and the coherent wMPFCA model is measured by the average RMSEs of ten sets of rolling-windows analysis. The minimal forecast errors are highlighted, and the lowest averaged forecast errors among models in different forecast horizons are highlighted in bold.} 
\label{table: Forecast accuracy of mortality for male and female by the average RMSEs in ten sets of rolling-windows analysis}
\end{table}
\end{landscape}

\begin{landscape}
\setlength{\belowcaptionskip}{1pt} 
\begin{table}[ht]
\ContinuedFloat
\begin{tabular}{cccccccccccccccc}  
\toprule
\multirow{2}{*}{
\parbox[c]{.06\linewidth}{\centering Country}}
& \multicolumn{3}{c}{Independent FPCA model} &&
\multicolumn{3}{c}{wMFPCA model} &&
\multicolumn{3}{c}{Product-Ratio model} && 
\multicolumn{3}{c}{Coherent wMFPCA model} \\ 
\cmidrule{2-4} \cmidrule{6-8} \cmidrule{10-12} \cmidrule{14-16}
& {\centering M} & {F} & {$\frac{\text{M+F}}{2}$} && {M} & {F} & {$\frac{\text{M+F}}{2}$} && {M} & {F} & {$\frac{\text{M+F}}{2}$} && {M} & {F} & {$\frac{\text{M+F}}{2}$}\\
\midrule
\underline{$h = 15$}\\
Australia & 0.2827  & \cellcolor{yellow}0.2289   & \cellcolor{yellow}0.2558 && 0.2959
          & 0.2780 & 0.2869 &&\cellcolor{yellow}0.2718 & 0.2476 & 0.2597 && 0.2982& 0.2793 & 0.2887\\
Belgium & \cellcolor{yellow}0.3132 & \cellcolor{yellow}0.3010 & \cellcolor{yellow}0.3071  && 0.4160 & 0.3917 &0.4039 && 0.3967 & 0.3685 & 0.3826 && 0.3516 & 0.3686& 0.3601\\
Canada & 0.2370   & \cellcolor{yellow}0.2040 & 0.2205  && 0.2513  & 0.2189 & 0.2351 && 0.2300 & 0.2105& 0.2203 && \cellcolor{yellow}0.2171   & 0.2133 & \cellcolor{yellow}0.2152 \\
France & 0.3096 & 0.2607 & 0.2852 &&0.3328& \cellcolor{yellow}0.2418 & 0.2873 && 0.2953 & 0.2730 & \cellcolor{yellow}0.2842 && \cellcolor{yellow}0.2887 & 0.2896 & 0.2892\\
Italy  & 0.4242 & \cellcolor{yellow}0.3776 & \cellcolor{yellow}0.4009&& 0.4923 & 0.4037 & 0.4480 && 0.5852 & 0.5477 & 0.5664 && \cellcolor{yellow}0.4013 & 0.4242   & 0.4127\\
Japan  & 0.2586  & 0.2426 & 0.2506 && 0.2394 & \cellcolor{yellow}0.2194 & 0.2294 && 0.2537 & 0.2330 & 0.2433 && \cellcolor{yellow}0.2075 & 0.2287 & \cellcolor{yellow}0.2181\\
Netherlands& 0.3229 & \cellcolor{yellow}0.2409 & 0.2819 &&0.4019  & 0.3739 &0.3879 && \cellcolor{yellow}0.2921& 0.2561
& \cellcolor{yellow}0.2741&& 0.2925 & 0.2602 & 0.2764 \\
Spain& 0.5495 & \cellcolor{yellow}0.3424 & 0.4459 && 0.6571 & 0.4969 & 0.5770 && 0.5982 & 0.5613 & 0.5798 && \cellcolor{yellow}0.4244 & 0.4360& \cellcolor{yellow}0.4302\\
U.K  & 0.2650  & \cellcolor{yellow}  0.2071& 0.2361 && 0.2837 & 0.2239 & 0.2538 && 0.2382& 0.2094& \cellcolor{yellow}0.2238 && \cellcolor{yellow}0.2291 & 0.2214 & 0.2252\\
U.S.A & \cellcolor{yellow}0.2054 & 0.1951  & 0.2002 && 0.2392 & 0.1798 & 0.2095  && 0.2133 & \cellcolor{yellow}0.1708
& 0.1921 && 0.2002  & 0.1877 & \cellcolor{yellow}0.1940 \\
Average  & 0.3168  & \cellcolor{yellow}\textbf{0.2600} & \cellcolor{yellow}\textbf{0.2884}   && 0.3609  & 0.3028  & 0.3319  && 0.3374 & 0.3078 & 0.3226 && \cellcolor{yellow}\textbf{0.2911} & 0.2909 & 0.2910 \\
\midrule
\underline{$h = 20$}\\
Australia    & 0.3449 & \cellcolor{yellow}0.2518& 0.2984   && 0.3418  & 0.3183  & 0.3301 && \cellcolor{yellow}0.3018  & 0.2693& \cellcolor{yellow}0.2855 && 0.3509 & 0.3307 & 0.3408 \\
Belgium   & \cellcolor{yellow}0.3430 & \cellcolor{yellow}0.3484 & \cellcolor{yellow}0.3457 && 0.4768& 0.4236& 0.4502&& 0.4186 & 0.4127 & 0.4156 && 0.3581 & 0.3998 & 0.3789\\
Canada  & 0.2708 & 0.2515& 0.2612 && 0.2824&0.2107& 0.2466 && 0.2688& 0.2657& 0.2672 && \cellcolor{yellow}0.2377
&\cellcolor{yellow}0.2340& \cellcolor{yellow}0.2358 \\
France &0.4459& 0.3317& 0.3888 && 0.3893 & \cellcolor{yellow}0.3239 & \cellcolor{yellow} 0.3566 && 0.3867 & 0.3551 & 0.3709 && \cellcolor{yellow}0.3769 & 0.4038 & 0.3903\\
Italy & 0.7292& 0.5598&0.6445&& 0.5603& \cellcolor{yellow}0.4528& 0.5065&& 0.6386& 0.5679& 0.6032&&  \cellcolor{yellow}0.4765& 0.5142 & \cellcolor{yellow}0.4953\\
Japan & 0.2906& 0.2702& 0.2804 && \cellcolor{yellow}0.2770& \cellcolor{yellow}0.2610& \cellcolor{yellow}0.2690&& 0.3004
& 0.3104& 0.3054&& 0.2865& 0.2818 & 0.2841\\
Netherlands  &0.3486& \cellcolor{yellow}0.2416& 0.2951&& 0.4621& 0.4607& 0.4614&& \cellcolor{yellow} 0.2875& 0.2536& \cellcolor{yellow} 0.2705&& 0.3023& 0.2637& 0.2830\\
Spain & 0.9573& \cellcolor{yellow}0.5407& 0.7490&& 0.9089& 0.6735& 0.7912&& 0.8918& 0.7700& 0.8309&& \cellcolor{yellow}0.6981& 0.6615& \cellcolor{yellow}0.6798\\
U.K &0.3709& \cellcolor{yellow}0.2484&0.3096&& 0.3909& 0.2902& 0.3406&& \cellcolor{yellow}0.2976& 0.2704 & \cellcolor{yellow}0.2840&& 0.3011& 0.2958& 0.2984\\
U.S.A & 0.2401& 0.2349& 0.2375&& 0.2750& 0.2038& 0.2394&&0.2590& \cellcolor{yellow}0.1950& 0.2270 && \cellcolor{yellow}0.2267& 0.2001 & \cellcolor{yellow}0.2134\\
Average& 0.4341& \cellcolor{yellow}\textbf{0.3279}& 0.3810&& 0.4365& 0.3619& 0.3992&& 0.4051& 0.3670& 0.3860&& \cellcolor{yellow}\textbf{0.3615} & 0.3585 & \cellcolor{yellow}\textbf{0.3600}\\
\bottomrule
\end{tabular}
\caption{(\textit{cont.}) Forecast accuracy of mortality for male and female using the independent FPCA model, the wMFPCA model, the Product-Ratio model and the coherent wMPFCA model is measured by the average RMSEs of ten sets of rolling-windows analysis. The minimal forecast errors are highlighted, and the lowest averaged forecast errors among models in different forecast horizons are highlighted in bold.} 
\end{table}
\end{landscape} 
\section{Discussion and conclusion remarks} \label{sec5: Discussion and conclusion remarks}
With the theoretical framework of multivariate functional principal component analysis motivated by \cite{chiou2016multivariate} and \cite{happ2018multivariate}, in this paper, we have proposed two new models that aim to model and forecast for a group of mortality rates, taking advantages of commonalities in their historical experience and age patterns. The first one, namely as wMFPCA model, is introduced to acknowledge differences in groups, age patterns and trends of several subpopulations to model together when subpopulations have somewhat sufficiently similar historical patterns. The coherent wMFPCA model is a novel extension of the wMFPCA model in a coherent direction. We design the coherent structure of the model to primarily fulfil the idea that when several subpopulation groups have similar socio-economic conditions or common biological characteristics and such these close connections are expected to continue and evolve in a non-diverging fashion in the distant future. The time weightings approaches on these two models lead us to expect the future patterns of mortality follow more likely recent past observations and obliterate some parts of irrelevant distant past mortality movements in favour of forecast performances of the two proposed models.
\par We have demonstrated the two models through forecasting for sex-specific mortality with the observed data from Japan. The wMFPCA model consists of the mean functions and the functional principal components of each subpopulation with corresponding scores shared by all subpopulations. We can obtain the forecasts of the wMFPCA model by extrapolating the shared principal component scores ahead with any non-stationary time series model, such as ARIMA model in our example. Whereas there are two primary components in the coherent wMFPCA model, including the average components among all subpopulations and the deviation components of each subpopulation deviated from the average components. We also use a non-stationary time series model for the predicted average component corresponding scores. However, we apply a stationary time series model for the forecast of deviation components corresponding scores so that the differences among all subpopulations will gradually decline and be bound to achieve coherence in the long term forecast. The forecasts of mortality sex ratios and life expectancy cure patterns confirmed the non-divergent forecastability of the proposed coherent wMFPCA model. 
\par The usefulness of the two proposed models is illustrated through a series of forecast accuracy evaluations and comparisons with other existing methods. The wMFPCA model provides a very flexible framework for multi-population mortality forecasting with comparable forecast accuracy as the independent FPCA model and the Product-Ratio model. The coherent wMFPCA model outperforms the Product-Ratio model in terms of forecast accuracy with no assumptions needed to place on the equal variance of all subpopulations. The coherent wMFPCA model maintains the same level in the short term forecast ability as the independent FPCA model. In contrast, the coherent wMFPCA model produces more sensible forecast results with less forecast error and bias in the two-sex mortality case in the long term.
\par The main limitations of the two proposed models also attribute to the characteristics in which they belong to the classes of `non-parametric' or `pure extrapolative' methods. They can capture trends in the historical data well. At the same time, they lack the ability to incorporate more other related information, such as the change in medical technology, environment and social-economy for predictions. Another issue is the compatibility of the wMFPCA models. It requires the sufficient level of homogeneity among subpopulations for forecasting, and the forecast ability of the wMFPCA models may be affected if several completely irrelevant subpopulations are placed together in the wMFPCA models. Although we used two sex-specific mortality rates for our demonstration in this article, it is obviously straightforward to apply these models to other scenarios with multiple related populations.

\begin{spacing}{1.5}
\bibliographystyle{apacite}
\bibliography{reference}
\end{spacing}
\end{document}